\documentclass[]{aa}
\usepackage{sidecap}
\usepackage{graphicx}
\usepackage[varg]{txfonts}
\usepackage{url}

\usepackage{gensymb}
\usepackage{natbib}
\bibpunct{(}{)}{;}{a}{}{,}
\usepackage{lscape}
\usepackage[dvipsnames]{xcolor}
\usepackage{amsmath}
\begin{document} 
\title{Constraining the physical structure of the circumstellar environment of V838 Monocerotis remnant}
   \author{Muhammad Zain Mobeen \inst{1}
          \and
          Tomasz Kami{\'n}ski\inst{1}
          \and
          Stephen Potter\inst{2,3}
}

\institute{\centering
Nicolaus Copernicus Astronomical Center, Polish Academy of Sciences, Toru{\'n}, Poland %\email{mzainmob@ncac.torun.pl}\label{inst1}
%\and
%\and
%Université Côte d’Azur, Observatoire de la Côte d’Azur, CNRS, Laboratoire Lagrange, France  \label{inst2}
%\and
%European southern Observatory, Karl-Schwarzschild-Str. 2, 85748 Garching bei Munchen, Germany\label{inst3}
%\and 
%European southern Observatory, Alonso de Cordoba 3107, Vitacura, Santiago, Chile\label{inst4}
%\and
%Astronomy Department, University of Michigan, Ann Arbor, MI 48109, USA\label{inst5}
%\and
%Astrophysics Group, Department of Physics \& Astronomy, University of Exeter, Stocker Road, Exeter, EX4 4QL, UK\label{inst6}
%\and
%Institut de Planetologie et d'Astrophysique de Grenoble, Grenoble 38058, France\label{inst7}
%\and
%The CHARA Array of Georgia State University, Mount Wilson Observatory, Mount Wilson, CA 91203, USA\label{inst8}
%NASA Ames Research Center, Moffett Field, CA 94035, USA \label{inst2}
\and
South African Astronomical Observatory, PO Box 9, Observatory 7935, Cape Town, South Africa \label{inst2}
\and
Department of Physics, University of Johannesburg, PO Box 524, Auckland Park 2006, South Africa \label{inst3}}

%\date{Received March, 2020}
\authorrunning{Mobeen et al.}
\titlerunning{Dusty enviroment of V838 Mon remnant}
% \abstract{}{}{}{}{} 
% 5 {} token are mandatory
 
\abstract{V838 Monocerotis (V838 Mon) erupted in 2002 as a luminous red nova after which it cooled and began to form dust. The remnant is predicted to become a blue straggler.  Interferometric observations in the $HKLM$ bands have uncovered a stable bipolar feature in the closest vicinity of the remant star.}{We aim to constrain the physical structure and nature of the circumstellar material immediately surrounding V838 Mon.}{Using the radiative-transfer code RADMC3D we managed to constrain the dust density distribution that best represents recent VLTI and CHARA interferometric imaging experiments. We also present recent SAAO-HIPPO polarimetric measurements which further test dust distibution models.}{We find that a multi-component model consisting of jets, a torus, and an ellipsoid provides an adequate fit to $H$ band interferometric observables, however, it struggles to reproduce the extremely small closure phase deviations in the $K$ band. The jets in our model are vital to produce nonzero closure phases, as they help to produce a gap in the torus which is the sole source of assymetry. The polarimetric measurements show that the intrinsic linear polarization is currently very low, with degree of polarization $<2\%$, consistent with our model.}{Although not unique, our dust model suggests a persistent torus-like structure and jet or jets in the remnant, in agreement with several predictions for an intermedite-mass merger product. However, even though these features may be important signposts of the evolution of V838 Mon to the blue straggler phase, the origin of these features remains largely unclear.}
%\vspace{2cm}

\maketitle

\section{Introduction}\label{intro}

  V838 Monocerotis erupted in a luminous red nova event \citep{2002A&A...389L..51M,2005A&A...436.1009T} in the beginning of 2002 and within a few weeks brightened by almost two orders of magnitude, reaching a peak luminosity of $10^{6} L_{\sun}$ \citep{2003Natur.422..405B,2005A&A...436.1009T,2008AJ....135..605S}. The event is believed to have been the result of a stellar merger. According to the scenario proposed in \cite{2006A&A...451..223T}, an 8 $M_{\sun}$ B-type main sequence star coalesced with a 0.4 $M_{\sun}$ young stellar object (YSO). The outburst was soon followed by a gradual decrease in temperature, and its spectra soon evolved to that of a late M-type supergiant \citep{2003MNRAS.343.1054E, 2015AJ....149...17L}. Spectral measurements in the 2000s revealed the presence of various molecules in V838 Mon, including water and transition-metal oxides \citep{ref161B,2009ApJS..182...33K}. Dust was also observed to be produced in the post-merger environment \citep{2008ApJ...683L.171W,alma}. Additionally, a B-type companion was observed at a distance of 225 au from the central merger remnant, which therefore suggests that the merger had taken place in a hierarchical triple system \citep{Bdiscovery2,alma}. The companion was obscured by dust formed in the aftermath of the 2002 eruption \citep{2009A&A...503..899T} and triggered dust formation in the passing merger ejecta \citep{alma}. V838 Mon is the most thoroughly observed luminous red nova in the Milky Way, although many others have also been found within the Galaxy as well as elsewhere in the Local Group \citep{2019A&A...630A..75P}. 
 
 Since the merger remnant in V838 Mon is enshrouded by dust, multiple mid-infrared (MIR) interferometric studies have targeted it. The first of these studies was conducted by \cite{2005ApJ...622L.137L}. They observed V838 Mon using the Palomar Testbed Interferometer (PTI) in 2004. By modeling the squared visibilities in the $K$-band at 2.2\,$\mu$m, they were able to measure the size of the dusty merger remnant to be $1.83 \pm 0.06$ mas. There were also hints of asymmetries in the object, but due to scarce measurements, these could not be confirmed. \cite{2014A&A...569L...3C} followed up on these measurements between 2011 and 2014, using the Very Large Telescope Interferometer (VLTI) instruments: Astronomical Multi-BEam combineR \citep[AMBER;][]{2007A&A...464....1P} in $H$ and $K$ bands, and the MID-infrared Interferometric instrument \citep[MIDI;][]{2003Ap&SS.286...73L} in the $N$ band. Uniform disk model fits to the AMBER measurements gave an angular diameter of $1.15 \pm 0.2$ mas, which, according to the authors, indicated that the photosphere of V838 Mon had contracted by about 40\% over the course of a decade \citep[this was later questioned in][]{2024A&A...686A.260M}. Also, \cite{2014A&A...569L...3C} modeling suggests that another extended component was present in the system, with a lower limit on the full width at half-maximum (FWHM) of $\approx$20\,mas. Modeling of the MIDI measurements  pointed towards the presence of a dusty elongated structure whose major axis varied as a function of wavelength between 25 and 70\,mas in the $N$ band. Furthermore, submillimeter observations obtained with the Atacama Large Millimeter/submillimeter Array (ALMA) in continuum revealed the presence of a flattened structure with a FWHM of 17.6$\times$17.6 mas surrounding V838 Mon \citep{alma}. Recent $L$ band measurements by \cite{2021A&A...655A.100M} paint a similar picture. \cite{2021A&A...655A.100M} geometrically modeled the squared visibilities and closure phases in the $L$-band, obtained using the Multi AperTure mid-Infrared SpectroScopic Experiment instrument (MATISSE) at the VLTI in 2020 \citep{2022A&A...659A.192L}. They found that the structure in the $L$-band is well represented by an elliptical uniform disk tilted at an angle of --40$\degr$. Furthermore, the closure phases showed small but non-zero deviations, which suggest the presence of asymmetries in the system. The interferometric measurements spanned across the wavebands (from 2.2 $\mu$m to 1.3 mm) and traced a dusty structure oriented roughly along the same direction, with a position angle (PA) in the range $-10$\degr\ (MIDI) to $-50$\degr\  (ALMA). This indicated either a single overarching structure in the post-merger remnant or multiple similarly aligned structures. Simulations of stellar merger events also suggest the presence of a disk-like structure in post-merger remnants, which is thought to be a reservoir for the pre-merger binary angular momentum \citep[e.g.][]{1976ApJ...209..829W, 2014ApJ...786...39N, 2017ApJ...850...59P, 2022ApJ...937...96M}. 

 In \cite{2024A&A...686A.260M} we were able to construct the first-ever milliarcsecond and sub-mas $H$ and $K$ band images using observations taken at the VLTI and CHARA arrays using the instruments GRAVITY/MATISSE and MYSTIC/MIRC-X, respectively. The size of the star in the $H$ and $K$ bands was $1.18$ mas ($\sim7$ au) and $1.94$ mas ($\sim14$ au), respectively. Using state-of-the-art image reconstruction packages, including SQUEEZE \citep{2010SPIE.7734E..2IB} and MIRA \citep{2008SPIE.7013E..1IT}, we attempted to reproduce the on-sky distribution of the circumstellar dust surrounding the merger remnant of V838 Mon. 
 %We were also able to perform speckle imaging of V838 Mon using the Zorro instrument in the visual range at Gemini South in 2021, which showed an extended elongated feature similar in orientation to what was reported for the NIR to MIR bands. 
 The resulting VLTI and CHARA images showed quite clearly a bipolar, possibly jet-like, feature in the immediate vicinity of V838 Mon. Bipolar outflows have been observed in a number of other red novae, suggesting that they can be quite common \citep{2018A&A...617A.129K,Kami2020,2024arXiv240103919K,2024A&A...682A.127S}, and their presence has long been advocated on theoretical grounds \citep{2021RAA....21...90S, 2024Galax..12...33S,Gagnier}. However, the coalesced star is very much similar now to red supergiants \citep{2020A&A...638A..17O,2023A&A...670A..13L} and, like those massive stars, it produces a dust-driven wind \citep{alma} whose erratic and clumpy nature can produce the reconstructed dust distribution as well. The nature of the circumstellar dust closest to V838 Mon remains thus unclear. Its understanding can prove to be important, as V838 Mon is currently the only known object that is a direct precursor of blue straggler stars \citep{2006A&A...451..223T,2024arXiv241010314W}. The mechanism and efficiency of losing angular momentum, be it through wind or jets, may explain characteristics of the different blue-straggler populations in young clusters \citep{2002ApJ...568..939L,2008MNRAS.384.1263C,2010AIPC.1314..105S}. 
 
While in the two previous papers we used geometrical modeling and image-reconstruction algorithms to reproduce the on-sky distribution of dust emission around V838 Mon,  in this study we attempt to go a step further by investigating whether we can place constraints on the three-dimensional dust distribution and on the dust physical properties. To this end, we construct and test models using the radiative transfer code RADMC3D \citep{radmc3d}. 
%
%Furthermore, we also study polarization changes that have occurred in the source over the last decade, and lastly, we also present recent Gemini observations in the $i$ band, this time obtained with the `Alopeke instrument. 
 %
 %V838 Mon serves as an excellent source to advance our understanding of the post-merger environment decades after the luminous red nova event, especially in the case of intermediate mass stars. Thus, it provides us with crucial insights into the physical processes at play in these merger events and their long-term evolution. In this paper, we analyze and interpret our recent interferometric observations obtained with a variety of instruments that span over NIR to MIR wavelengths. 
%
We mainly focus here on observables delivered by the VLTI and CHARA interferometers, along with complementary polarization measurements. All these observations trace the immediate circumstellar material that is located very close to the central merger remnant, within 10--300 au, as opposed to the large-scale (900 au) roughly spherical ejecta seen through millimeter CO emission by \cite{alma}. % The distance to V838 Mon is relatively well known \citep[5.9 kpc;][]{2020A&A...638A..17O}, at which 10 mas corresponds to 59 au.

% The format of the paper is as follows. In Sect. \ref{Observations} we present all of our VLTI and CHARA observations and outline the main steps of the data reduction. We also analyse and interpret recent optical speckle interferometric observations obtained at 562 nm and 832 nm. In Sect. \ref{geomod} we mainly present the results of geometrically modeling the interferometric observables (squared visibilities and closure phases) observed with the MATISSE and GRAVITY instruments at VLTI and with MIRCX/MYSTIC at CHARA. Sect \ref{imaging} centers around our image reconstruction attempts for the VLTI and CHARA datasets using two distinct image reconstruction algorithms. The modeling and imaging results are discussed in depth in Sect. \ref{disc} finally followed by Sect. \ref{concl} in which we present the main conclusions of this study.     
%%%%%%%%%%%%%%%%%%%%%%%%%%%%%%%%%%%%%%%%%%%%%%%%%%%%%%%%%%%%%%%%%%%%%%%%%%%%%%%%%%%%%%%%%%%%%%%%%%%%%%%%%%%%%%%%%%%%%%%%%%%%%%%%%%%%
\section{Observations}\label{Observations}
%{\bf SHORTRECAP OF THE INTERF. OBSERVATIONS HERE}.
\subsection{Recap of interferometric observations}
This paper is mainly based on interferometric observations presented in \cite{2024A&A...686A.260M}.
The VLTI observations were carried out in 2022, using the 1.8 m Auxiliary Telescopes (ATs) with the GRAVITY instrument. GRAVITY operates in the $K$ band. Multiple configurations were used for the GRAVITY observing blocks, with baseline lengths of 30--140 m. These included the large, small, and medium arrays along with intermediate extended configurations. % in order to increase $uv$ plane sampling for image reconstruction. 

V838 Mon was observed at CHARA in March 2022 using the MIRC-X/MYSTIC combined instrument mode. MIRC-X operates in the $H$ band, while MYSTIC operates in the $K$ band. The goal for these observations was also imaging, which is why we made use of the extended five-telescope configuration. The CHARA baselines span the range 34--330 m, thus the longest CHARA baselines is three times longer than the longest VLTI baseline. This makes sub-mas imaging possible with CHARA. Since the CHARA MYSTIC $K$ band data was taken mainly for fringe tracking and not imaging, in \cite{2024A&A...686A.260M} we relied exclusively on GRAVITY $K$ band for image reconstruction. Here we also omit the MYSTIC data. Furthermore, since we were unable to obtain extensive imaging observations with MATISSE, we could not construct images in the $LM$ bands. They are not considered here. The MIRC-X and GRAVITY interferometric observations are, however, complemented by original and and nearly contemporary polarimetric measurements presented below. 

\subsection{SAAO-HIPPO}\label{sect-obs-pol}   
The polarimetric observations were made with the HIPPO polarimeter \citep{2008SPIE.7014E..5EP} on the South African Astronomical Observatory (SAAO) 1.9 m telescope. The $VRI$ observations were obtained on 7 October 2021 and the $B$ filter observations on 9 October 2021, i.e., a few months before the interferometric measurements. The unpolarized standard star HD176425 was observed to verify instrumental polarization, and the polarized standard star HD187929 was observed to verify the instrument zero points and polarization efficiencies. These were then used to calibrate the measurements of V838 Mon. Data reduction was carried out as described in \cite{2010MNRAS.402.1161P}. Both linear and circular polarization measurements were obtained. No circular polarization was detected within the uncertainties of the instrument, giving upper limits of $\approx$0.1\% in the $VRI$ bands and 0.5\% in the $B$ band. The measured total polarizations along with the computed intrinsic polarizations are given in Table \ref{HIPPO_pol_2021}. The magnitudes in the $BVRI$ bands during this epoch were $15.5$, $13.2$, $11.0$, and $8.9$, respectively\footnote{\url{http://www.vgoranskij.net/v838mon.ne3}}. V838 Mon is significantly fainter in $B$, resulting in much larger polarization uncertainties.

To extract the intrinsic polarization, i.e., the polarization inherent to V838 Mon, we removed the interstellar medium (ISM) polarization component from the observed total polarization. Our correction procedure is described in Appendix \ref{appendix-polarization} and is based on ISM polarization parameters derived from observations of field stars in \citet{wisniewski2003b}. The results are presented in Table \ref{hippo-pols}. The intrinsic position angle (PA) values vary considerably between the bands. For example, while the $V$ and $R$ band position angles are fairly similar (i.e., $-17\degr$ and $-11\degr$, respectively), the $I$ band position angle ($\approx78\degr$) is vastly different. Owing to relatively low polarization levels, these individual measurements have large uncertainties; the results are in agreement within these large errorbars. The position angle averaged over all the bands is $16\fdg3 \pm 39\fdg9$. Note that the total polarization is only marginally higher than the ISM polarization and considering realistically all uncertainties (e.g., systematic errors in deriving the ISM polarization from field stars), we may be getting a null polarization signal from V838 Mon.

\begin{table*}%[hb!]
\caption{Polarization results for our SAAO-HIPPO observations in 2021. Model results are presented in the last column.}
\label{hippo-pols}
\centering
%\hspace*{-1cm}
\begin{tabular}{l c c cc  c|c}
\hline\hline
Band & $P_{T}$ & $PA$ & $P_{\rm ISM}$ & $P_{int}$ & $PA_{int}$ &$P_{model}$  \\ 
& [\%] &$^\circ$& [\%]&[\%]& $^\circ$ & [\%] \\
\hline
%[0.5ex] % inserts table %heading
$B$  &$4.105 \pm 1.333$  &$170.584 \pm 11.866$ & $2.582 \pm 0.013$ & $1.807 \pm 2.841$ &$15.55 \pm 30.26$ & $2$ \\
$V$  &$3.824 \pm 0.229$ & $155.963 \pm 2.422$ & $2.738 \pm 0.011$ & $1.095 \pm 0.471$ & $-17.61 \pm 8.52$ & $1.6$\\
$R$  &$3.095 \pm 0.102$ &$155.371 \pm 1.324$ &$2.710 \pm 0.012$  & $0.397 \pm 0.498$ & $-11.053 \pm 10.66$ & $0.6$\\
$I$  &$2.419 \pm 0.091$ &$158.807 \pm 1.522$ &$2.469 \pm 0.014$ &$0.236 \pm 0.767$ & $78.37 \pm 22.16$ & $1$\\%[5pt]
\hline
\end{tabular}
\label{HIPPO_pol_2021}
\tablefoot{$P_{T}$, $PA$, $P_{\rm ISM}$, $P_{int}$, $PA_{int}$ and $P_{model}$ are the total polarization, position angle, ISM polarization, intrinsic polarization, the position angle of the intrinsic polarization, and the average linear polariazation of our RADMC3D model respectively. $P_{\rm ISM}$ was calculated for the observed wavelengths from the Serkowski's law fit in \citet{wisniewski2003b} and PA of 153\fdg43 was adopted for the ISM component after \citet{wisniewski2003b}.}
\end{table*}

%%%%%%%%%%%%%%%%%%%%%%%%%%%%%%%%%%%%%%%%%%%%%%%%%%%%%%%%%%%%%%%%%%%%%%%%%%%%%%%%%%%%%%%%%%%%%%%%%%%%%%%%%%%%%%%%%%%%%%%%%%%%%%%%%%
\section{Radiative transfer modeling with RADMC3D} \label{geomod}
\label{analysis}
The primary aim of this study was to identify and characterize the dusty environment of the V838 Mon remnant by constructing a physical model that would explain the recent NIR interferometric imaging. 

The most relevant data products for the above-mentioned VLTI and CHARA instruments are the interferometric observables, i.e., squared visibilities ($V^{2}$) and the closure phases (CPs). The squared visibilities represent the fringe contrast.  An object is said to be completely resolved in the case that the value for the squared visibility is zero and completely unresolved in the case that the value is one. The squared visibilities can be used to constrain the size of the source. Closure phases are the sum of the individual phase measurements by telescopes within a particular triangular configuration in the array. This results in the atmospheric phase cancelling out. The phase of the complex visibility function is sensitive to the object symmetry \citep{1958MNRAS.118..276J}. Closure phases are a probe for asymmetries, so deviations from values of 0\degr\ or 180\degr\ would indicate some deviation from centro-symmetry of the source at the spatial scales probed by the VLTI and CHARA baselines. 

%\subsection{RADMC3D}\label{L-model}

In our modelling, we utilize RADMC3D \citep{2012ascl.soft02015D}, which is a radiative transfer code that allows the user to create their own custom dust distributions and define the radiation sources. Dust opacity files are defined based on the grain composition and size distributions. The code computes dust temperatures under the assumption of radiative equilibrium. Using ray tracing and Monte Carlo Markov Chain (MCMC) algorithms, RADMC3D yields synthetic images, which can then be directly compared with observations. 

%\footnote{we used radmc3dPy \url{https://www.ita.uni-heidelberg.de/~dullemond/software/radmc-3d/manual_rmcpy/radmc3dPy.html} for this purpose.}. 

%%%%%%%%%%%%%%%%%%%%%%%%%%%%%%%%%%%%%%%%%%%%%%%%

We set up a 3D grid comprising 250 $\times$ 250 $\times$ 250 points and used a single star as the sole source of radiation located at the center of the grid (the companion of the remnant is well outside the considered volume). The stellar temperature was set to 3500 K, which was derived from the observed spectra of V838 Mon \citep[cf.][]{alma}. In RADMC, the stellar spectrum is approximated by a Planck function for this temperature, while the luminosity of $10^{5}L_{\odot}$ is set in the program by the stellar radius of 1420 $R_{\odot}$. For all our simulations presented in this section, we used opacities for amorphous silicates composed of 70\% Mg-rich and 30\% Fe-rich pyroxene from \citet{opacity1} and \cite{opacity2}. As V838 Mon is oxygen-rich and displayed the silicate spectral feature just after the 2002 outburst \citep{rushton2005infall}, this generic opacity function is an adequate first guess and was used  for modeling the spectral energy distribution of the remnant in the past \citep{alma}. The opacity file represents dust grains with the canonical `MRN' size distributions \citep{MRN}.

\begin{figure}
    \centering
    \includegraphics[trim= 170 135 100 155, clip, width=\columnwidth]{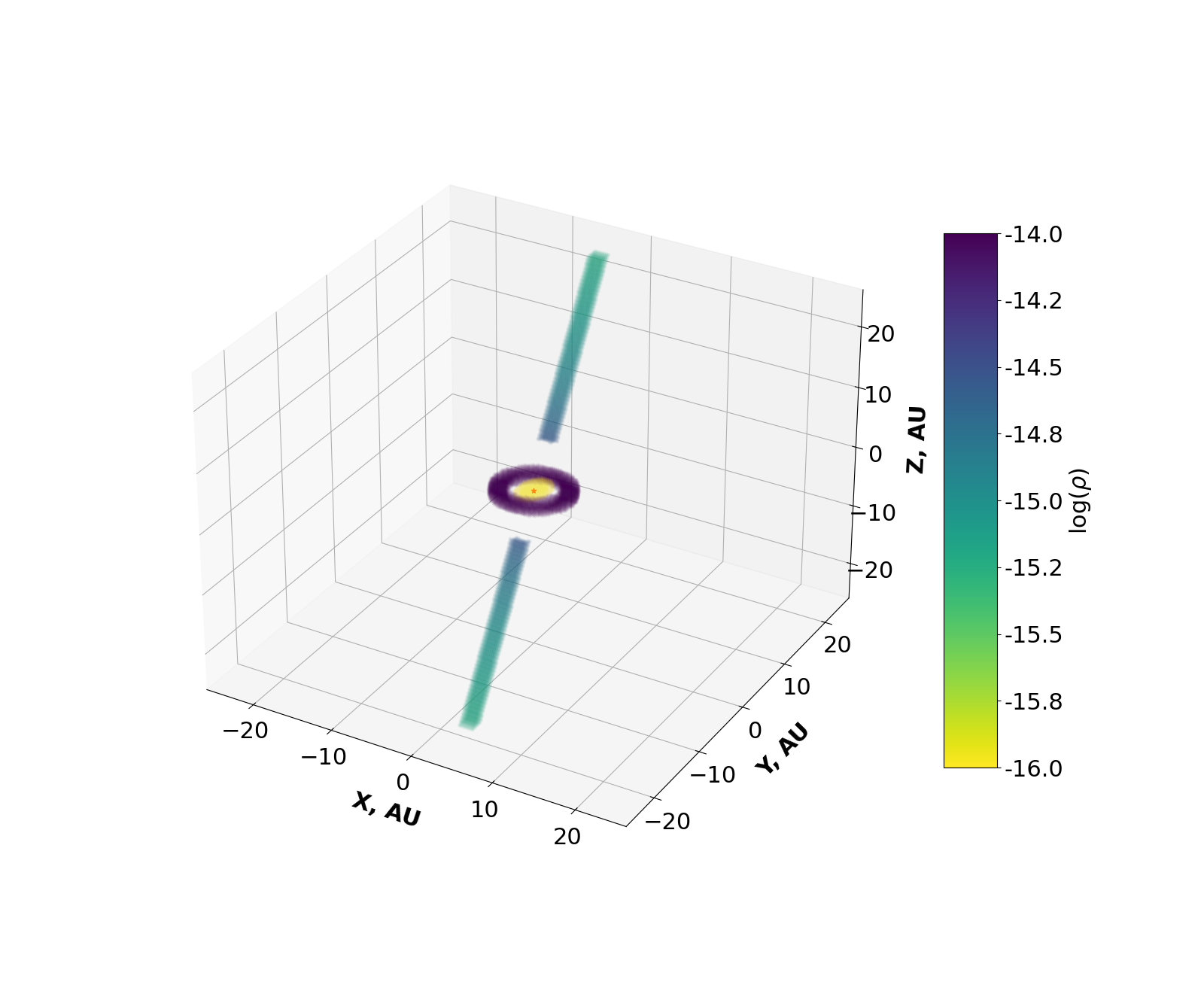}
    \caption{Density structure of the modeled components. The 3D view shows the system at an arbitrary projection (different than the observer's view point) to show all the components. The star position is marked with the red star symbol. The density, $\rho$, is in g cm$^{-3}$.}
    \label{fig3d}
\end{figure}

In constructing the first models, we took a cue from the image reconstructions presented in \cite{2024A&A...686A.260M}. At the distance of V838 Mon (i.e., $\sim$ 5.9 kpc), 1 mas corresponds to a physical scale of $\sim$ 6 au. In \cite{2024A&A...686A.260M}, the extent of the circumstellar dust traced in the $HK$ bands is $1-2$ mas. which is why the dust distribution in the model was restricted to about 6 au, so that we could reproduce realistic synthetic images in the $HK$ bands. In \cite{2024A&A...686A.260M}, image reconstructions suggested the presence of a bipolar feature in V838 Mon. Previously, \cite{alma} also found an extended elongated dusty structure in the sub-mm regime, which is oriented along the same position angle as that of the feature found by \cite{2024A&A...686A.260M}. This is in congruence with what is observed in other red novae, such as V4332 Sgr \citep{2018A&A...617A.129K}, V1309 Sco \citep{2024A&A...682A.127S}, CK Vul \citep{kaminski2021molecular}, or in the merger candidate the Blue Ring Nebula \citep{2020Natur.587..387H}. Furthermore, numerical simulations \citep{kashi2009ngc,pejcha2017pre,2020ApJ...893...20S,2023arXiv230607702S} seem to indicate that the formation of jet-like outflows are associated with merger events and their aftermath. Thus, we molded our modelling approach in light of the aforementioned observational and theoretical results. The model that we present and test in this study, as well as the additional models shown in the appendix, is comprised of three distinct components. 1) an inner ellipsoid that immediately surrounds the central star, 2) a slightly extended tilted toroidal feature, and 3) a pair of collimated jets perpendicular to the plane of the torus and extending well beyond it. This configuration was thoroughly tested and adjusted to best reproduce the observables. The model is presented in Fig. \ref{fig3d}.

Initially, we considered models without the inner central ellipsoid that directly surrounds the star. However, in such models, the point source representing the star would completely dominate the emission and leave no appreciable circumstellar structure that is readily present in the interferometric observations. To rectify this problem, we surrounded the star with the dusty ellipsoid, which resulted in more diffusion of the stellar light. In our final model, the diameters of the inner ellipse are 3 au $\times$ 2 au $\times$ 0.5 au. The jets and the inner ellipsoidal feature are oriented along a position angle of $20\degree$ measured anti-clockwise from north, while the torus is oriented at $-60\degr$. The ellipsoid is thus thinnest along the line of sight.

The torus is the most massive and most dense component of the model. Through trial and error, the inner and outer radii of the torus were set to 3 au and 5 au, respectively (see Fig.\ref{fig3d}). 

The jets were made to begin from 7 au and extend out to nearly the edge of the modeled domain. The jets have intermediate density among the three components considered here but are not prominent in the simulated images. This is mostly owing to the relatively large distance of the outermost parts of the jets from the radiation source. Increasing their density results in higher optical depths and self-shadowing. It is thus difficult to produce bright linear jets within the adopted framework. 

The density of the inner ellipse and the torus was set to $10^{-14}$ g cm$^{-3}$. The density profile of the jet was set to $2\times10^{-14}$ g cm$^{-3}/r$, where $r$ is the distance of the jet from the star in au. The resulting total dust mass we considered in most simulations was $\approx 10^{-6}\,M_{\odot}$. This is three orders of magnitude less than the dust mass obtained from ALMA observations \citep{alma} of the much cooler dust directly surrounding V838 Mon (i.e., not accounting for the more extended merger ejecta). The part of the circumstellar matter bright at NIR wavelengths is indeed expected to constitute only a small fraction of the overall dust surrounding V838 Mon. We probe here the warmest and innermost dust of the post-merger remnant. The density structure of the modeled components is shown in Fig. \ref{fig3d}.

While the different components do not overlap with each other spatially (Fig. \ref{fig3d}), in projection one line of sight may cross all three of these features. In consequence, for instance, the northern jet, which is closer to the observer than the torus, absorbs some of the radiation from the torus. That can result in a gap in the brightness distribution of the torus, like in Fig. \ref{bestmodH} (north-southern gap).
 
In our initial RADMC3D runs, we also performed a sublimation temperature correction for the dust grains. Any dust with an equilibrium temperature exceeding the sublimation temperature, somewhat arbitrarily set to 1700 K (see below), was removed from the modeled volume. We intended to apply this correction iteratively until a steady configuration is reached. This, however, turned out to remove nearly all dust. At the high luminosity of the star and the limited optical depth of the dust, the radiation effectively sublimates any dust out to the outermost radii considered here. This is not consistent with observations at IR and mm wavelengths, as dust is readily present within a few au of the star. Increasing the sublimation temperature to higher values considered in the literature for silicate dust does not alleviate this problem (not even 2700 K postulated for the most refractory alumina dust). We also considered other forms of dust, but most species considered in the literature heat up even more effectively than our silicate mixture. It is possible that the considered dust analogs are not a good representation of the real dust surrounding V838 Mon. Even more likely, the environment may be constantly producing and destroying dust, which is impossible to model within the static RADMC3D framework. We therefore did not apply the automatic sublimation correction. The hottest dust in our models has a maximum temperature of $\sim7000$ K but its mass is very low compared to the total dust mass considered in our RADMC3D simulations. The presence of the sublimating dust means that the dusty environment of V838 Mon is unlikely to be a long-lived structure and may be undergoing rapid changes all the time.

%to requirement to destroy the dust entirety of the dust distribution would be cleared away, leaving no viable and stable feature in the circumstellar vicinity. We then also considered other dust species, in order to see if there exists an alternative with lower opacity in the $HK$ bands. The best species was the silicate mixture that was already being used in our models. Thus, we modified our approach by removing the sublimation temperature correction, as a result of which we were able to create various models. We discuss the implications and nature of such physical models which are formed from extremely hot dust in Sect.

\begin{figure}%[hbt!]
    \centering
    \includegraphics[trim=15 5 35 35,clip, width=\columnwidth]{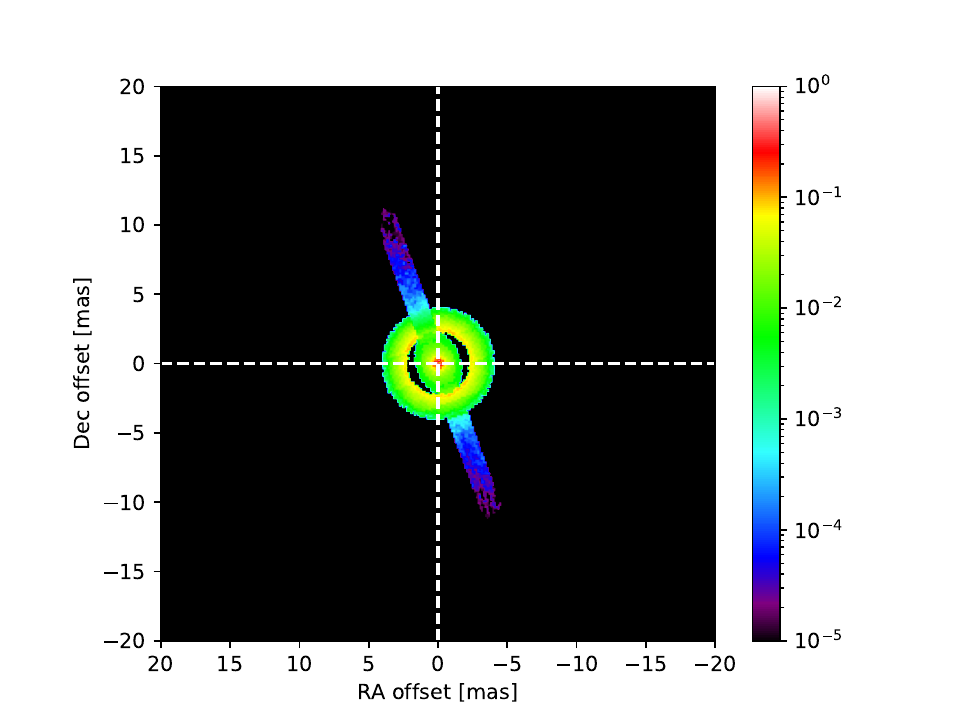}
    \caption{$K$ band simulated image for our best model. The colorbar shows normalized logarithmic intensity. Note that the dynamic range limit of the array is of $\approx$100 which means parts of the displayed structure, e.g., the jets, would not be registered by the interferometers.}
    \label{bestmodK}
\end{figure}

\begin{figure}%[hbt!]
    \centering
    \includegraphics[trim=15 5 35 35,clip, width=\columnwidth]{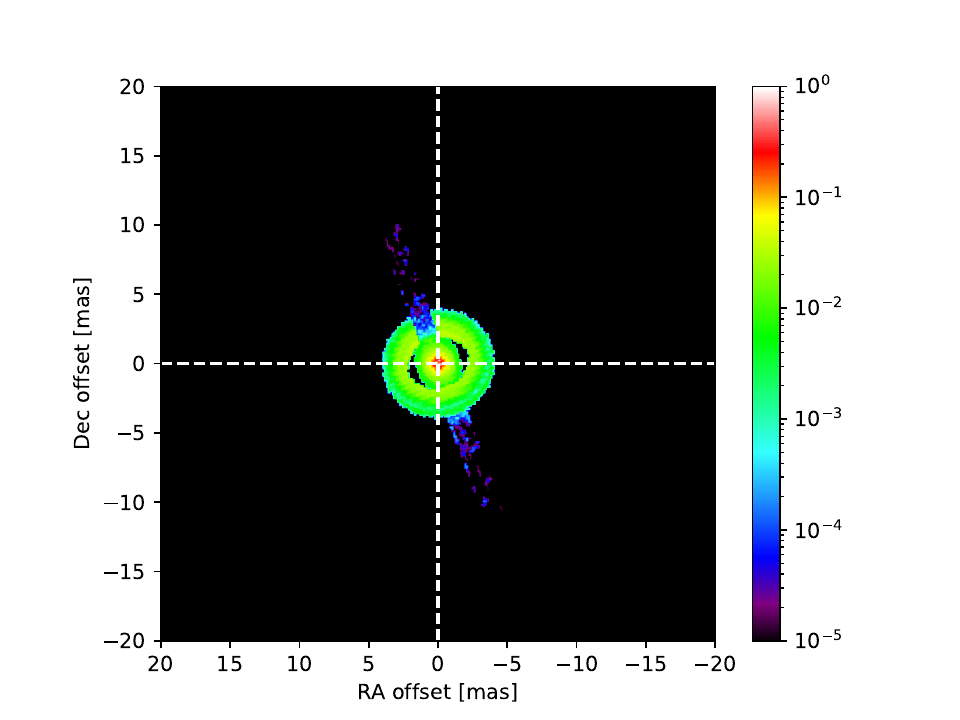}
    \caption{$H$ band simulated image for our best model. The colorbar shows normalized logarithmic intensity. }
    \label{bestmodH}
\end{figure}

Once we completed the calculation of thermal equilibrium temperature in RADMC3D, we generated on-sky images in the $H$ and $K$ bands. Simulated images are constructed using a ray tracing method. For visualizations, we utilized the radmc3dPy package\footnote{\url{https://www.ita.uni-heidelberg.de/~dullemond/software/radmc-3d/manual_rmcpy/radmc3dPy.html}} with the function {\tt makeImage}. We specified the number of pixels to 300. % and chose 10 points for the wavelength ranges. 
The number of photon packages tracked in MCMC was adjusted to minimize simulation artifacts. The size of the image was restricted to 50 au, which corresponds to an angular size of $\sim$7 mas. The simulated images for our preferred model can be seen in Figs. \ref{bestmodK} and \ref{bestmodH}.

While we present here only one specific family of models with a torus and jets, we also examined several other {\it ad hoc} configurations of clumps and other structures. However they did not show promise of representing the observations well, especially the closure phases. Some of these models are briefly described in online materials\footnote{\url{https://doi.org/10.5281/zenodo.15316096}}.
%https://zenodo.org/records/15316097?token=eyJhbGciOiJIUzUxMiJ9.eyJpZCI6IjIxYzljNmNkLWFlZjEtNDI1MC1iZWU0LTM1NzJmN2FjMjdlOCIsImRhdGEiOnt9LCJyYW5kb20iOiIwOGY5ZDUzMDE5N2YyN2Y4MjFjNTdhNWUwYTA4OTQ1MiJ9.z0DxkHbBUGHkPssdX0_oTZo5DsGfBGKJZoGiEMbnasAcnvmmb5G_RL2qIECWob4QEfOi5suLWlva4sF37N7Deg}}.

%%%%%%%%%%%%%%%%%%%%%%%%%%%%%%%%%%%%%%%%%%%%%%%%%%%%
\subsection{Optimizing the model to the interferometric observables} \label{results}
The $H$ and $K$ images obtained from RADMC3D were used to generate synthetic squared visibilities and closure phases that could be compared directly to their observed counterparts from \cite{2024A&A...686A.260M}. In order to compute these quantities, we used the JMMC software AMHRA\footnote{\url{https://amhra.oca.eu/AMHRA/oifits-modeler/input.htm}}. As inputs, AMHRA takes (i) the OIFits files \citep{OIfits} with the observations (i.e., in our case, the VLTI and CHARA data for the $K$ and $H$ bands, respectively) and (ii) the FITS files generated in radmc3dPy representing the physical model in consideration (see Figs. \ref{bestmodK} and \ref{bestmodH}). The input OIFits observation file ensures that the simulated observables are identical in baseline and wavelength to the observations. 
%The user can then choose the observables to be computed. For the purpose of this study, we computed the squared visibilities and closure phases only, since those could be directly compared to the observations. The resulting output OFITS file contains these two observables.
%
%\subsection{Interferometric Observables from RADMC3D}
%\label{results}
%As mentioned above, we were able to obtain simulated squared visibilities and closure phases for each of the six models described above. 
These simulated observables were then compared to the observed squared visibilities and closure phases, as shown in Figs. \ref{bestmodKvis} and \ref{bestmodHvis}. In the following sections, we discuss the results for the presented model in both bands.
%Our main motivation for this modeling exercise was to see if the peculiarities of the observations could be replicated with any of the physically motivated models. 

\subsection{$K$ band}

\begin{figure}%[hbt!]
    \centering
    \includegraphics[trim=10 5 40 36,clip, width=\columnwidth]{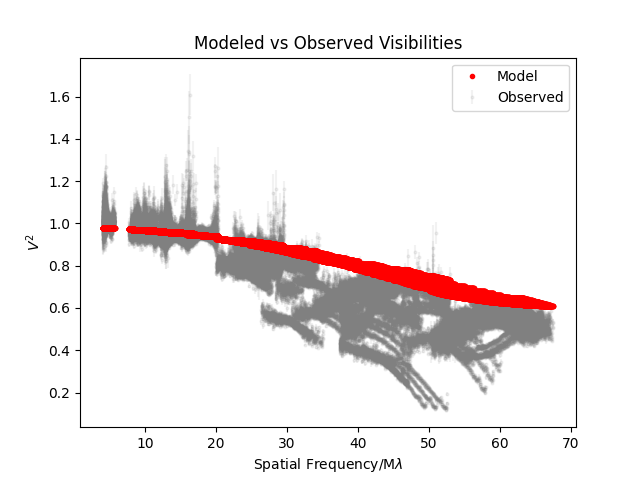}
    \includegraphics[trim=10 5 40 36,clip, width=\columnwidth]{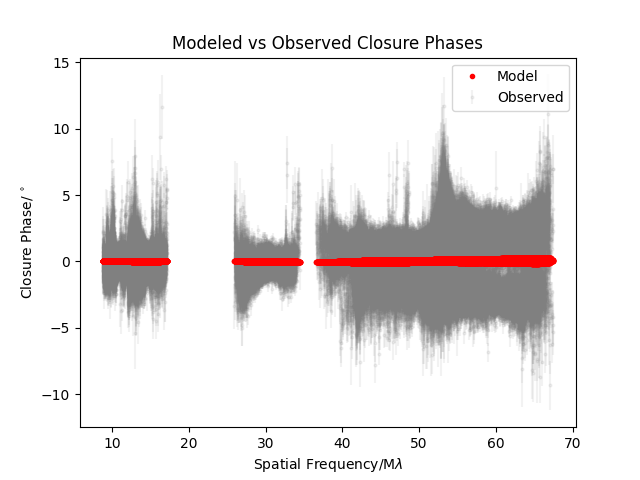}
    \caption{$K$ band simulated (red) and observed (grey) visibilities. The top panel shows squared visibilities, while the bottom panel shows closure phases.}
    \label{bestmodKvis}
\end{figure}

\begin{figure}%[hbt!]
    \centering
    \includegraphics[trim=10 5 30 36,clip, width=\columnwidth]{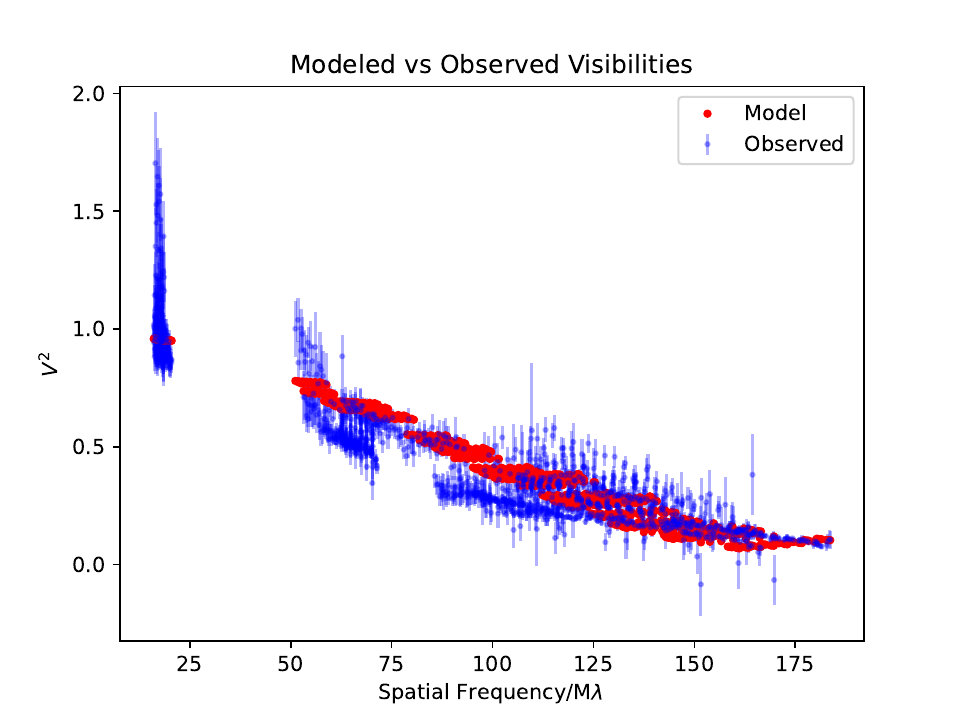}
        \includegraphics[trim=10 5 30 36,clip, width=\columnwidth]{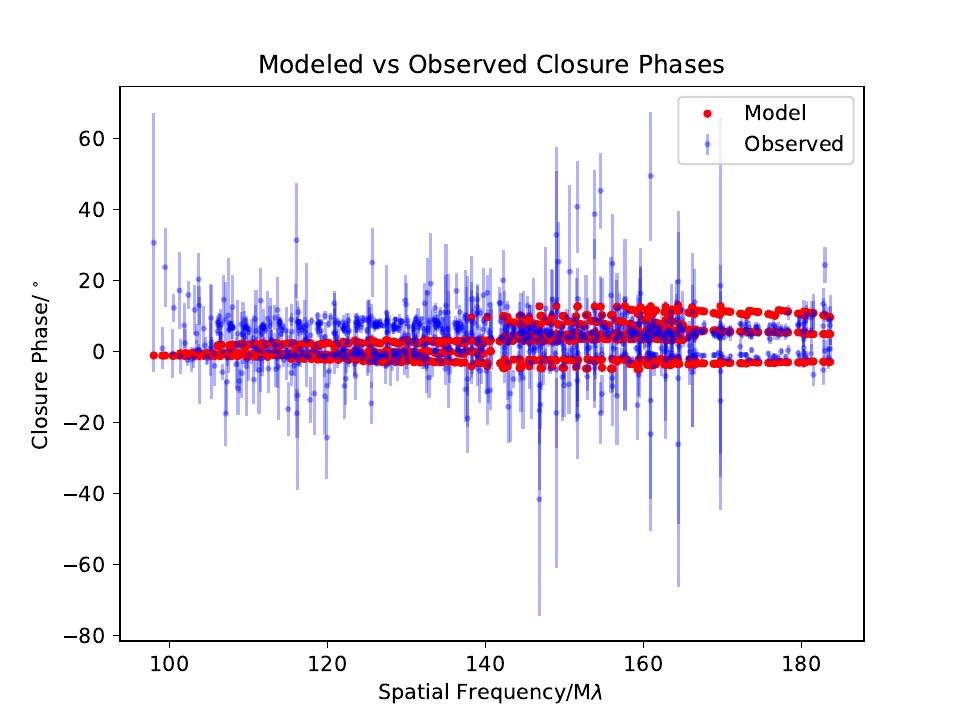}
    \caption{$H$ band simulated (red) and observed (blue) visibilities. The top panel shows squared visibilities, while the bottom panel shows closure phases.}
    \label{bestmodHvis}
\end{figure}

The simulated $K$ band squared visibilities show a gradual decline in value with increasing spatial frequency. They reach a minimum of 0.6 at a spatial frequency of $67 M\lambda$. This means that the model image in the $K$ band remains unresolved even at the longest baselines, since $\sim 40\%$ of the flux remains unexplained. Another feature of these simulated visibilities is the noticeable absence of the bump-like features that were reported in \cite{2024A&A...686A.260M}. This can be seen in Fig. \ref{bestmodKvis}. 

The simulated $K$ band closure phases show almost no deviations from zero. At higher spatial frequencies, there is a very slight increase in closure phase to $\sim$ 0.2\degr\ (see Fig. \ref{bestmodKvis}). This is about an order of magnitude smaller than the closure phase scatter seen in the observations. Our chosen model is unable to fully reproduce the observed closure phase signal observed with GRAVITY in $K$ band. 

%\begin{figure}%[hbt!]
%    \centering
    
%    \caption{$K$ band simulated and observed closure phases, plotted as a function of spatial frequency. Blue points are observations, while red points represent the model closure phases.}
%    \label{bestmodKt3}
%\end{figure}

We carried out a number of other simulations in order to see if we can get simulated $K$ band visibilities less than 0.6 at higher spatial frequencies, since the observed $K$ band visibilities do tend towards zero at higher spatial frequencies. These models can be seen in Appendix \ref{modplots} and were created by expanding the outer radius of the torus component. We explored various values of outer radii from 8 to 40 au (see Figs. \ref{3to8image}--\ref{3to40image}). We found that for an outer radius of 40 au, we obtained a minimum squared visibility of 0.4 (Fig. \ref{3to40vis2}). Beyond this value of outer radius, the squared visibility at longer spatial frequencies did not decrease. Our modelling efforts have revealed that it was not possible to explain the remaining $40\%$ of the unresolved flux. With regard to the various sizes that we tested, we note that these lie in the angular size range of $1.5-7$ mas. %The $K$ band image reconstructions from \cite{2024A&A...686A.260M} place an upper bound of $\sim 2$ mas on the size of the structure. Given this size constraint, any model with a physical size greater than 6 au is not feasible and therefore cannot be given any credence. 

In Fig. \ref{bestmodK} the jets are the least prominent component of the three, with an intensity of $\sim 10^{-5}$, well below the GRAVITY  dynamic range limit. We note that none of the components in the $K$ band model are able to reproduce the closure phase signal. This means that the jets, along with the other two components, are not major sources of asymmetry in the $K$ band.   

We conclude that our model qualitatively reproduces observations quite well in the $K$ band but does not capture intricate asymmetries that produce quick declines of $V^2$ at some baselines (observation position angles). This would mean that in the $K$ band the circumstellar environment is a lot more complex than the models that we have assumed in this study. 

\subsection{$H$ band }
\label{hbandanalysis}
\begin{figure}%[hbt!]
    \centering
    \includegraphics[trim=10 5 30 36,clip, width=\columnwidth]{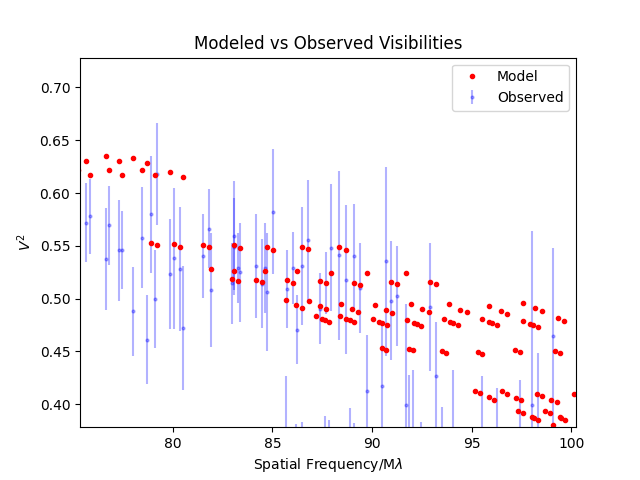}
    \caption{A zoomed-in section of the observed vs model squared visibilities plot in Fig. \ref{bestmodHvis}.}
    \label{bestmodHvis_zoom}
\end{figure}

%The difficulty we encountered in accurately reproducing the $H$ band closure phases could be due to some peculiarity that maybe unique to this band. This would mean that in the $H$ band the circumstellar environment is a lot more complex than the models that we have assumed in this study.    
The simulated squared visibilities in the $H$ band generally follow the gradual downward trend with increasing spatial frequency, as seen in the observations. The simulated data at the highest spatial frequency reach values of $<$0.1, which means that our model image is in fact well resolved by the CHARA baselines used for the observations. Another trend seen is the plateauing of individual clusters of visibilities with increasing baseline length (see Fig. \ref{bestmodHvis_zoom}). Beyond a spatial frequency of 100M$\lambda$, within these clusters, there is a slight upward turn in the squared visibilities. This latter trend is in opposition to the behavior of the visibilities seen in the spatial frequency range $60-90$ M$\lambda$, where the individual clusters show a clear downward-sloping trend. At a spatial frequency of around $60 M\lambda$, the observed visibilities also show a bifurcation due to which there is a smaller cluster of squared visibilities in the spatial frequency range $60-70$ M$\lambda$, with points clearly lower than the rest of the measurements in that range. A similar bifurcation also occurs across the range $90-120$ M$\lambda$. Our model is unable to account for this subtle feature. 

%\begin{figure}%[hbt!]
%    \centering

%    \caption{$H$ band simulated and observed squared closure phases, plotted as a function of spatial frequency. Blue points are observations, while red points represent the model closure phases.}
%    \label{bestmodHt3}
%\end{figure}

The observed closure phases in $H$ band show a clear demarcation of two populations of measurements, one of which is centered at zero degrees while the other is at around $10^{\circ}$. There are also a number of closure phase points greater than $15^{\circ}$, hinting towards significant asymmetries in the source. Our model closure phases in the $H$ band show clear deviations from $0^{\circ}$ at higher spatial frequencies (i.e., $>$ 140 M$\lambda$), some of them even touching $\sim$ $13^{\circ}$. We are able to somewhat reproduce the observed cluster at $0^{\circ}$ as well as half of the second cluster centered at about $10^{\circ}$. The observed closure phase measurements seen in the range $110-140$ M$\lambda$ are not reproduced in the simulated data. The model is successful at reproducing accurately the closure phase points at higher spatial frequencies.  

In order to see if a model could be constructed that successfully explains the bifurcation of the visibilities in the $H$ band, we experimented with various additional models (Appendix \ref{appendix-H}). Nearly all the models yielded squared visibilities that were much worse than the main model that we present. There was one model (Fig. \ref{additionalHband}) in which we were able to reproduce the lower groups of visibility points in the ranges where the bifurcation occurs (Fig. \ref{additionalHvis2}). In this model the position angle of the jets and the torus has been changed to $-20\degree$, while the inner ellipsoid now lies along a position angle of $20\degree$. Also, the ellipsoid in this particular model has been compressed along its intermediate axis by a factor of 4. The squared visibilities from this model are underestimated over the rest of the spatial frequency range. Furthermore, the closure phases (Fig. \ref{additionalHt3}) for this model show distinct peaks in both directions at higher spatial frequencies. This is not at all visible in the observations, which is another factor for not choosing this model. 

In the $H$ band we find that the jet component in our model helps to explain the noticeable closure phase deviations. Without including a jet in the models, we noticed that the resulting simulated closure phases were $<1\degree$. In Fig. \ref{bestmodH} we can see that at the base of the jet where it overlaps in projection with the torus, there is a small gap. This gap produces an asymmetric intensity distribution in the torus.  This is in stark contrast to the $K$ band models in which including the jets does not yield significant closure phases since the torus remains unaffected. %The jet component in the $H$ band models is, therefore, of great significance and may be hinting towards interactions between outflows and any other dusty features in the circumstellar vicinity in V838 Mon. 

\subsection{Model orientation}

We also constructed a number of other dust models where we experimented with various orientations. This was done by changing certain parameters when using the {\tt makeImage} function in radmc3dPy, those being the polar inclination angle, $i_{obs}$, and the azimuthal angle, $\phi_{obs}$. The definitions and visual illustrations of these angles in the image plane can be seen in the online radmc3dPy manual\footnote{\url{https://www.ita.uni-heidelberg.de/~dullemond/software/radmc-3d/manual_radmc3d/imagesspectra.html}}. The model images for the various values of  $i_{obs}$ and $\phi_{obs}$ can be seen in Appendix \ref{appendix-angles} (Figs. \ref{incl50}--\ref{incl90phi50}). The goal of this exercise was to see if changing the orientation could produce better simulated squared visibilities and closure phases in the $H$ band. In particular, we wanted to test if the bifurcation in the visibilities, along with the closure phase signal at shorter spatial frequencies, can be reproduced. The results are shown in Figs. \ref{incl50vis2}--\ref{incl90phi20t3}. It is clear that none of these inclined models can reproduce both observables accurately. For example, in the case of the squared visibilities, all models produce an extra cluster of visibilities between 100--150 M$\lambda$ that is not seen in the observations (Figs. \ref{incl50vis2}--\ref{incl90phi20vis2}). The closure phases are almost always underestimated, except for the $i_{obs} = 70\degree$ model, in which the deviations at higher spatial frequencies are extremely large, i.e. $>100\degree$ (Fig. \ref{incl70t3}). This led us to exclude these models from our analysis.

%\subsection{Polarization in V838 Mon}
\subsection{Polarization RADMC3D setup}
Using RADMC3D we were able to generate polarized-light images for our best model. To fully account for scattering, we used the RADMC3D default opacity files for the classic silicate dust, which contain all the necessary optical properties to run the polarization models with full scattering. This dust is composed of amorphous silicate grains of a size of 0.1 $\mu$m and containing equal amounts of Fe and Mg-rich olivines from \citet{1994A&A...292..641J} and \citet{opacity2}. The opacity curves of this silicate mixture are only slightly different than the mixture of silicates we used in the modeling described in the previous sections. The full scattering mode of RADMC3D allows for the complete treatment of polarized light scattering off randomly oriented particles. In this particular mode, the code uses the full dust scattering matrix instead of just using the Henyey-Greenstein anisotropic phase function. This allows for taking into consideration the effect of the viewing angle as well\footnote{\url{https://www.ita.uni-heidelberg.de/~dullemond/software/radmc-3d/manual_radmc3d/dustradtrans.html#sec-scattering}}. Calculating the polarization maps required 900 million photon packages and was more computationally costly than the simulations discussed so far.

Using the full Stokes' parameters $IQU$ calculated by the program, we generated maps presenting the degree of linear polarization (DoLP). Sample maps, in $V$ and $I$ photometric bands, are shown in Fig. \ref{VIbandpol} (extra maps for the $K$ and $H$ bands, and for $V$ with adjusted scaling, are shown in Appendix \ref{appendix-polarization-maps}). The DoLP was computed for each pixel as $\sqrt{U^{2} + Q^{2}}/I$. We then calculated an intensity-weighted mean over the DoLP maps (i.e., $\sum DoLP\cdot I/\sum I$), which gave us the total degree of linear polarization for V838 Mon (i.e. $P_{model}$) in $BVRI$ bands used in the HIPPO observations. These values are given in the final column of Table \ref{HIPPO_pol_2021} and are all lower than 2\%. They are similar to the SAAO-HIPPO intrinsic polarization given in Table \ref{HIPPO_pol_2021}.  Our model can accurately reproduce the observed polarimetric measurements. The average polarization in $HK$ bands is $\sim1\%$.

As shown in Fig. \ref{polmapsboth}, the DoLP is greatly skewed towards larger values (i.e. $>50\%$) in the jet regions, whereas in the torus the DoLP remains relatively low (i.e. $<2\%$). The total polarization $P_{model}$, given its low values, is therefore dominated by the polarized emission arising from the torus, which is brighter.

\begin{figure}%[hbt!]
    \centering
    \includegraphics[trim=10 5 0 15,clip, width=\columnwidth]{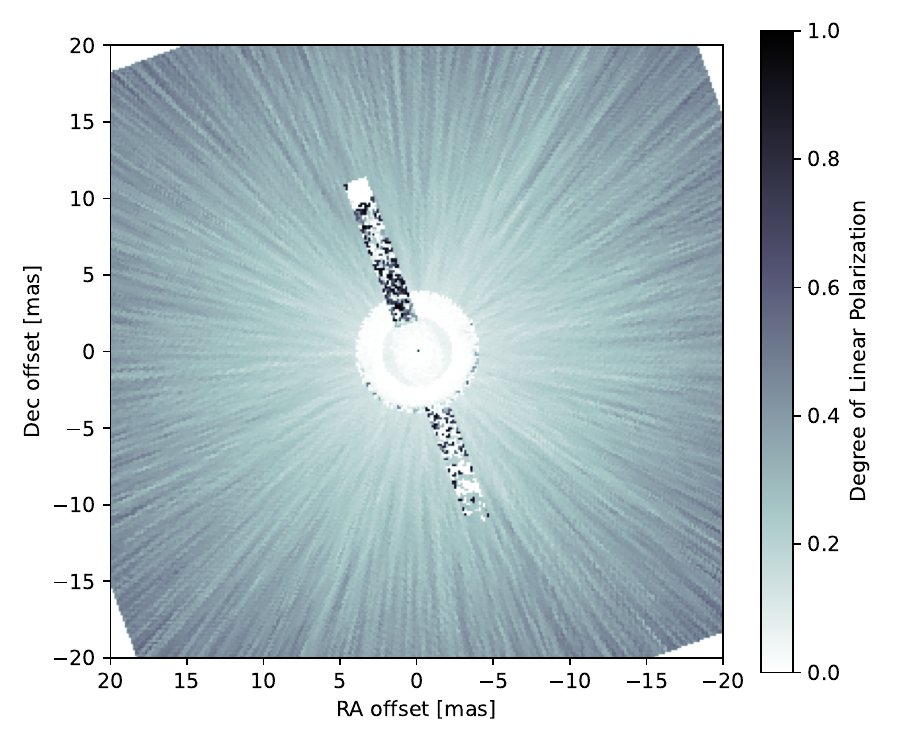}
%    \caption{$K$ band simulated image with degree of linear polarization.}
%    \label{kbandpol}
%\end{figure}
%\begin{figure}%[hbt!]
%    \centering
    \includegraphics[trim=10 5 0 15,clip, width=\columnwidth]{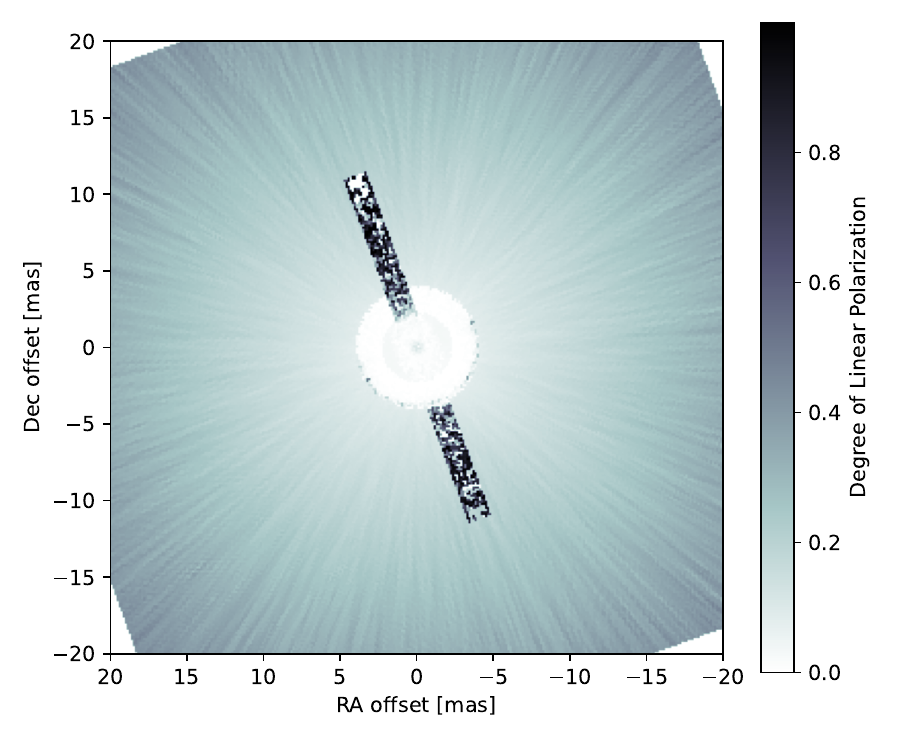}
    \caption{Simulated maps of degree of linear polarization in $V$ (top) and $I$ (bottom) bands. Concentric rays are due to imperfect ray tracing.}
    \label{VIbandpol}
\end{figure}

%We find that in the case of the $HK$ bands, the average polarization is $\sim1\%$. Our model therefore predicts very low polarization in the $HK$ bands. While in the $BVRI$ bands, the average degree of polarization is $<2\%$ (see last column in Table \ref{HIPPO_pol_2021}). These values are similar to the SAAO-HIPPO values for the intrinsic polarization given in Table \ref{HIPPO_pol_2021}, which suggests that the model can accurately reproduce the observed polarimetric measurements.   

%This is roughly similar to the degree of polarization seen in the recent SAAO-HIPPO observations for the $BV$ bands, displayed in Table \ref{HIPPO_pol_2021}. In the images, we can see that some very large values of degree of polarization (i.e. $>80\%$) are present in the jets, but the surface area and surface brightness of these regions is insignificant.  

%

\subsection{SiO masers}
\label{maser}

Maser emission which has been observed in V838 Mon, may help constrain the physical parameters of the nearest circumstellar environment of the merger remnant. SiO maser emission is generally found around red supergiants (RSGs), asymptotic giant branch (AGB) and post-AGB stars with stable outflows. %SiO maser emission is used to determine the size of the photosphere for such objects since the maser spots are located in the immediate vicinity of the radio-photosphere, which typically is twice as large as the optical--IR stellar disk \citep{grey}. 
SiO masers were also observed in the disk wind of Orion KL's SrcI \citep{2010ApJ...708...80M, 2017A&A...606A.126I}, a suspected merger remnant of a young system \citep{Bally}. There, the masers are seen above and below the plane of a dusty disk surrounding the protostar(s).

Under the assumption of collisional pumping, these SiO masers occur at molecular hydrogen densities of $10^{8} - 10^{10}$ cm$^{-3}$. For radiative pumping, they require the presence of a dusty envelope \citep{2000PASJ...52..895M,2017A&A...606A.126I}. The low-frequency SiO masers near 43 and 86 GHz, which have been observed in V838 Mon since the end of the eruption \citep{deguchi2005detection,claussen,2020A&A...638A..17O}, specifically require excitation of the $\nu = 1,2$ vibrational levels of SiO. Only luminous stars with mass loss rates higher than $10^{-5}$ M$_{\sun}$ yr$^{-1}$ are observed to have such SiO maser emission \citep{2016ApJ...817..115C}.

\begin{figure}%[hbt!]
    \centering
    \includegraphics[trim=75 5 50 35, clip, width=\columnwidth]{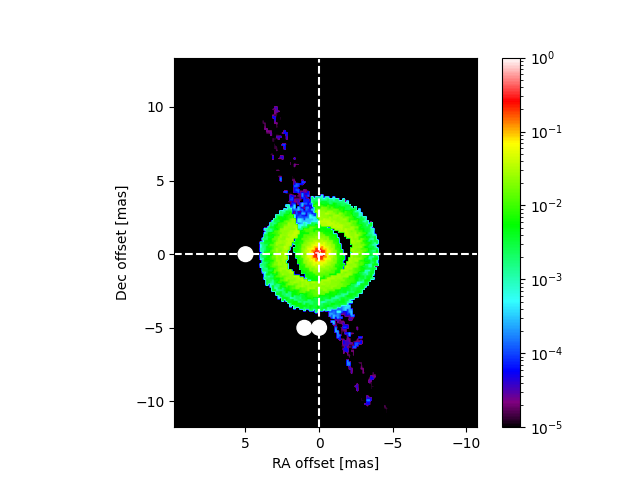}
    \caption{Location of SiO maser spots relative to the $H$ band image of our best dust model. The three white circles represent the three strongest spots detected by \cite{2020A&A...638A..17O} with the VLBA at 43 GHz.  We located the spots with respect to the brightest pixel, assuming this is the real position of the star.}
    \label{maserplots}
\end{figure}

\cite{2020A&A...638A..17O} reported the positions and proper motions of the SiO $\nu = 1,2$ and $J=1-0$ and $2-1$ maser transitions near 43 GHz around V838 Mon. They identified three main spots located in the vicinity of the stellar remnant in V838 Mon. We marked the positions of these maser spots on the $H$ band image for our best model in Fig. \ref{maserplots}. The two brightest spots are separated by $\approx 6$ mas. Characteristically, the masers are seen only south and east relative to V838 Mon and are just outside the continuum emission of our model torus (for some models considered in Appendix \ref{appendix-K}, they are at the edge or within the torus). Note that SiO masers in V838 Mon, like in SrcI, are not overlapping with bright continuum in the best model. Such an asymmetric distribution of maser spots has often been observed around genuine red supergiants \citep[e.g.,][]{2013MNRAS.436.1708R,2025arXiv250305250S} and is often assigned to stochastic shocks propagating through the envelope \citep{2014A&A...572L...9R,grey}.  Following the model of \cite{grey}, the dust number densities should be $10^{-4} - 10^{-3}$ cm$^{-3}$, which -- adopting an average grain radius of 0.05 $\mu$m and density of 3 g cm$^{-3}$ -- puts the requirement on dust density of $10^{-17} - 10^{-16}$ g cm$^{-3}$. This is below the density of about 10$^{-14}$ g cm$^{-3}$ implemented for the torus in our models (cf. Fig. \ref{fig3d}).

The presence of the SiO masers indicates that the NIR interferometric observations probe a region around V838 Mon that is very dynamic and affected by shocks. Our best model is not inconsistent with the location of the maser spots but the complex circumstances of SiO maser action do not allow us to put firm constraints on the physical properties of the medium.

\section{Discussion}
\label{disc_sec}

Our proposed model is in fair agreement with interferometric and polarization observations. While this model is not unique and there may be other configurations that  explain the observations equally well, or better, our simulations show that it is possible that a torus and jets are present in the post-merger system of V838 Mon. Below, we briefly discuss what it may imply.

\subsection{Jets and outflows in a merger remnant}
\label{jetsandoutflows}

In Sect. \ref{hbandanalysis}, we mentioned that the closure phase deviations are only explained with the gap in the torus that results because of the jet. Our modelling efforts prove that both components are necessary for understanding the closure phases at higher spatial frequencies. The orientation of the torus matches well with the general orientation of the circumstellar material seen in other bands \citep{alma,2021A&A...655A.100M}. Meanwhile, the jets are almost perpendicular to the torus, though they are not bright enough to be detected with either CHARA or VLTI. However, jets are thought to be a key mechanism for explaining the evolution of physical properties in luminous red novae (LRN). Simulation results by \cite{2025PASP..137c4201S} propose a mechanism in which accretion onto a main sequence star can cause the removal of mass from the outer layers of the star via the launching of jets from the accretion disk. Previously, \cite{2023arXiv230607702S} also found that the outbursts observed during an LRN event are difficult to explain without the launching of jets. In the case of V838 Mon they claim that an accretion disk was formed around the primary 8 $M_{\odot}$ star due to the disruption of the YSO, and then jets were launched from the accretion disk. It is possible that the torus in our model represents the remnant of the accretion disk that was formed around the primary star in the initial stages of the merger. In the two decades since the outburst, the jets might have become considerably less bright. \cite{2019MNRAS.483.5020S} propose that the jets formed in LRN can interact with the surrounding low- density medium, causing the formation of bubbles and clumps via Rayleigh-Taylor instabilities. Some kind of dynamical change must have occurred in the post merger environment that produces the observed asymmetry. 

It is also likely that whatever the nature of this change, it could be occurring rapidly.  We mentioned in Sect. \ref{analysis} that dust temperatures in our simulations were quite high, which suggests that the components in our model are likely transient in nature. Interestingly, the variable nature of LRN jets was suggested by simulations performed by \cite{2023ApJ...954..143D}. The authors found that jets formed during the common envelope phase, as a result of accretion of gas from RSGs onto a companion compact object, will 'wobble' with respect to the angular momentum vector. The change in the angle of the jets is estimated to be in between $10-40\degr$. In the case of V838 Mon, though, it is not well established whether the merger involved the common-envelope evolution. However, the initial angular momentum of the system could have been deposited in jets after the merger, causing them to precess. If indeed a jet of this type does persist in V838 Mon then a possible line of investigation could be to perform long-term multi-band polarimetry of V838 Mon. The 2021 SAAO-HIPPO results indicate that the polarization position angles in the $V$ and $R$ bands have changed drastically by $\sim 100 \degree$ since the previous measurements by \cite{wisniewski2003b} in 2003. A frequent and long-term polarimetry campaign could help to answer the question of jet variability definitively.

Recently \cite{ibrahim2023imaging} found using CHARA that the Herbig Be star HD 190073 is surrounded by a disk-like feature which shows temporal variations in its intensity distribution. In particular, the images over a span of a few months reveal a bright feature that seems to be rotating at sub-Keplerian speeds. The observed closure phases with MIRC-X are not too dissimilar from the ones we present in this paper, with their closure phases reaching up to $\sim20\degree$ at higher spatial frequencies. The authors suggest multiple phenomena that could produce the observed asymmetry in HD 190073 such as planet formation, spots or self-shadowing of the disk due to its inclination. These seem unlikely in the case of V838 Mon. Changing the viewing angle for our model was unable to yield adequate closure phases, which rules out any inclination-related effects. Given the high dust temperatures of the model, it seems likely that, similar to the disk in HD 190073, the $H$ band structure could also be varying on short timescales of $\sim$ a few months. Follow-up CHARA observations can help to further constrain the nature of the structure in the $H$ band. For example, if subsequent interferometric observations show a very clear change in the closure phase signal such as a boost or a drop, then such a change can be attributed to dynamical changes occurring within either the torus or the jets. The 2022 observations show a clear preference for the model with the jets inclined along a North East position angle, i.e. $20\degree$. A significant deviation from this angle is something that the $H$ band closure phases as measured by CHARA will be sensitive to, as evidenced by the additional model that we presented in Sect. \ref{hbandanalysis} (see Figs. \ref{additionalHband}--\ref{additionalHt3}). Future CHARA/VLTI observations can be used to establish a baseline to gauge the frequency and the magnitude (in terms of position angle change) of the ensuing variations in $HK$ band post merger environment.

\subsection{Features of evolution into a blue straggler?}
V838 Mon, having a mass of about 8 $M_{\odot}$ and being a cluster member, is currently undergoing a relaxation phase which could result in the formation a blue straggler. How blue stragglers are formed, especially how those originating in mergers get rid of angular momentum, is still debated \citep{2010AIPC.1314..105S,2024arXiv241010314W}. Instead of interpreting the modeled dust structures as the direct aftermath of the merger, we can alternatively view them as signposts of the ongoing transformation of V838 Mon. The innermost structure, located well within the torus, is most likely dusty wind which at the physical parameters of V838 Mon must be intrinsically variable. It is interesting to ask whether it may provide the star with a way to get rid of excess angular momentum predicted for merger products. Similarly, the jets and torus may be a product of an enhanced rotation \citep[cf.][]{2006A&A...453.1059K} and rotation-induced excretion, or even enhanced magnetic fields. Investigating these components deeper has the potential of addressing many interesting problems related to blue stragglers' formation.

\subsection{V838 Mon: An ongoing story}

This study is the third in a series of publications that has sought to answer the pressing question with regard to the nature of the circumstellar environment in V838 Mon. In \cite{2021A&A...655A.100M} for the very first time, using MATISSE observations at VLTI, we showed that the interferometric observables suggested the presence of a disk like component along with very minute asymmetries in the $L$ band. We then sought to fully image the dusty environment in V838 Mon. Subsequently, in \cite{2024A&A...686A.260M} we made use of observations taken using MIRC-X/CHARA and GRAVITY/VLTI. This allowed us to construct high resolution images which clearly showed a bipolar morphology in the $H$ and $K$ bands. This is in accordance with what has been observed in other merger objects such as CK Vulpeculae \citep{2018A&A...617A.129K} and the Blue Ring Nebula \citep{2020Natur.587..387H}. The bipolar lobes in V838 Mon also appeared to be slightly asymmetric, especially in the $H$ band. This led us to speculate that the lobes could be transient and may be undergoing dynamical changes over small timescales. The RADMC3D model that we present in this paper seems to be indicating that V838 Mon is composed of two main components, the torus and the jets. The jet component, despite its faintness, is crucial in order to explain the observed asymmetry in the system. The high dust temperatures hint at a potentially transient feature in the $H$ band. This warrants further observational studies using interferometry and polarimetry as discussed in Sect. \ref{jetsandoutflows}. This will enable us to answer the question of variability in V838 Mon decisively, and the results can be used to further improve theoretical models of LRN outbursts and their aftermath.

\section{Conclusions} \label{concl}

In this paper we presented RADMC-3D models that fairly well explain the $H$ and $K$ band observables obtained at the VLTI (GRAVITY) and CHARA (MIRC-X). They also explain the most recent 2021 polarimetric data  measured using the HIPPO instrument at SAAO.

%\begin{itemize}

%\item  
The RADMC3D modeling in the $H$ band suggests a multi-component model comprising jets, a torus and an inner ellipsoid surrounding the central star In the $K$ band all attempted models are degenerate, and the squared visibilities are underestimated.

%\item 
In the model the jets interact with the torus to produce a gap in the torus which helps to explain the observed closure phase signal. The presence of this gap in the models raises many interesting questions about the potentially transient nature of the post-merger environment in the $H$ band.

%\item The $VR$ polarization measurements from 2021 show that a huge shift ($\sim 100\degree$) in the PA has occurred in V838 Mon since the measurement made by \cite{wisniewski2003b}.  

%\end{itemize}

\bibliographystyle{aa}
\bibliography{export-bibtex.bib}

\begin{acknowledgements}
T. K. and M. Z. M. acknowledge funding from grant no 2018/30/E/ST9/00398 from the Polish National Science Center. Based on observations made with ESO telescopes at Paranal observatory under program IDs 0104.D-0101(C) and 0108.D-0628(D). This research has benefited from the help of SUV, the VLTI user support service of the Jean-Marie Mariotti Center (\url{http://www.jmmc.fr/suv.htm}). This research has also made use of the JMMC's  Searchcal, LITpro, OIFitsExplorer and Aspro services available at \url{http://www.jmmc.fr/}. This research has made use of the Jean-Marie Mariotti Center, JSDC catalogue available at \url{http://www.jmmc.fr/catalogue_jsdc.htm}. This research has made use of the Jean-Marie Mariotti Center (JMMC) — MOIO AMHRA service at https://amhra.jmmc.fr. We would also like to thank Noam Soker for going over this manuscript and providing us with valuable feedback.

\end{acknowledgements}

\begin{appendix}

\section{Intrinsic polarization and position angle calculation}\label{appendix-polarization}

The intrinsic polarization is given by the following relation

\begin{equation}
    P_{int} = \sqrt{P_{T}^2(\sin{{\theta}_{obs}} + \cos{{\theta}_{obs}})^2 - P^{2}_{\rm ISM}(\cos{{\theta}_{\rm ISM}} + \sin{{\theta}_{\rm ISM}})^2}.
\end{equation}
$P_{T}$ and $P_{\rm ISM}$ are the observed total polarization and the ISM polarization, respectively; $\theta_{obs}$ and $\theta_{\rm ISM}$ are the position angle for the total polarization vector and the position angle for the ISM polarization, respectively. We computed the intrinsic polarization angle, $\theta_{int}$, as
\begin{equation}
%    \theta_{int} = \arctan{\frac{P_{T}\cos({180\degr- \theta_{obs}}) - P_{\rm ISM}\cos({180\degr-\theta_{\rm ISM}})}{P_{T}\sin({180\degr-\theta_{obs}}) - P_{\rm ISM}\sin({180\degr-\theta_{\rm ISM}})}}.
    \theta_{int} = \arctan \frac{P_{T}\cos\theta_{obs}-P_{\rm ISM}\cos\theta_{\rm ISM}}{P_{T}\sin\theta_{obs}- P_{\rm ISM}\sin\theta_{\rm ISM}}.
\label{pa_eq}
\end{equation}
The position angles are measured positive anti-clockwise from north. The ISM polarization was calculated using Serkowski's law
\begin{equation}
    P_{\rm ISM} = P_{max}\exp\Bigg({-K \log \bigg( \frac{\lambda_{max}}{\lambda}}\bigg)^2\Bigg).
\label{serkowskieq}
\end{equation}   
%\section{$K$ band closure phases} \label{Kcps}
Here $P_{max} = 2.746 \pm 0.011$, $\lambda_{max} = 579 \pm 4$ nm, $K = 0.971$, $\theta_{\rm ISM}$=153\fdg43. These values were taken directly from \cite{wisniewski2003b}, who derived them from observations of field stars. The $BVRI$ band centers for our observations were set at 450, 550, 650, and 806 nm, respectively. 

The errors on the polarization values and position angles were obtained using standard error propagation methods. The total error was calculated using the error in the ISM polarization. This error was then propagated along with the errors in other quantities. The expression for the error in the ISM polarization ($\delta P_{\rm ISM}$) is given by
\begin{equation}
\begin{split}
\sigma_{P_{\text{ISM}}} = P_{\text{ISM}} \Bigg[  
\left( \frac{\sigma_{P_{\max}}}{P_{\max}} \right)^2 +  
\left( \frac{2 K}{\lambda_{\max}} \log \left( \frac{\lambda_{\max}}{\lambda} \right) \sigma_{\lambda_{\max}} \right)^2 + \\
\left( \frac{2 K}{\lambda} \log \left( \frac{\lambda_{\max}}{\lambda} \right) \sigma_{\lambda} \right)^2  
\Bigg]^{\frac{1}{2}}
\end{split}
\end{equation}

The error formula for the intrinsic polarization (i.e. $\delta P_{int}$) is

\begin{equation}
\begin{split}
\sigma_{P_{\text{int}}} = \frac{1}{P_{\text{int}}} \Bigg[
P_T^2 (\sin\theta_{\text{obs}} + \cos\theta_{\text{obs}})^4 \sigma_{P_T}^2 + \\
P_{\text{ISM}}^2 (\sin\theta_{\text{ISM}} + \cos\theta_{\text{ISM}})^4 \sigma_{P_{\text{ISM}}}^2 + \\
P_T^4 (\sin\theta_{\text{obs}} + \cos\theta_{\text{obs}})^2 (\cos\theta_{\text{obs}} - \sin\theta_{\text{obs}})^2 \sigma_{\theta_{\text{obs}}}^2 + \\
P_{\text{ISM}}^4 (\sin\theta_{\text{ISM}} + \cos\theta_{\text{ISM}})^2 (\sin\theta_{\text{ISM}} - \cos\theta_{\text{ISM}})^2 \sigma_{\theta_{\text{ISM}}}^2
\Bigg]^{\frac{1}{2}}.
\end{split}
\end{equation}

%\begin{figure*}%[hbt!]
%    \centering
%    \includegraphics[clip, width=18cm, height=11cm]{T3PHI_15.pdf}
%    \caption{Closure phases versus wavelength in the $K$ band.}
%    \label{kband_t3_15}
%\end{figure*} 

\section{Additional $K$ band models}\label{appendix-K}
\label{modplots}

\begin{figure}%[hbt!]
    \centering
    \includegraphics[clip, width=\columnwidth]{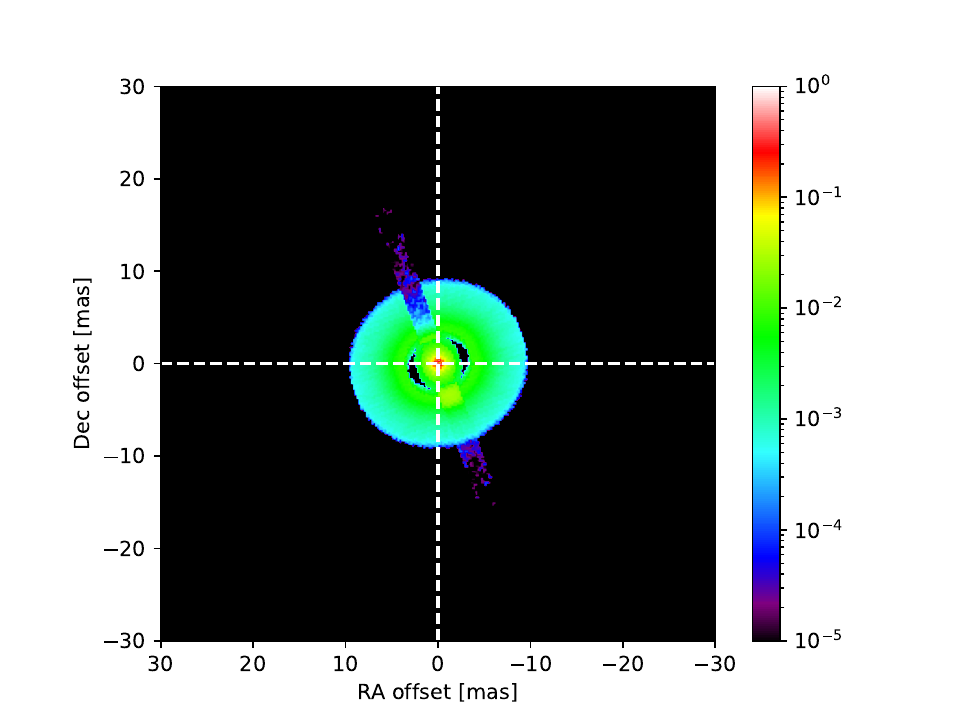}
    \caption{$K$ band model image with a torus outer radius of 8 au.}
    \label{3to8image}
\end{figure}

\begin{figure}%[hbt!]
    \centering
    \includegraphics[clip, width=\columnwidth]{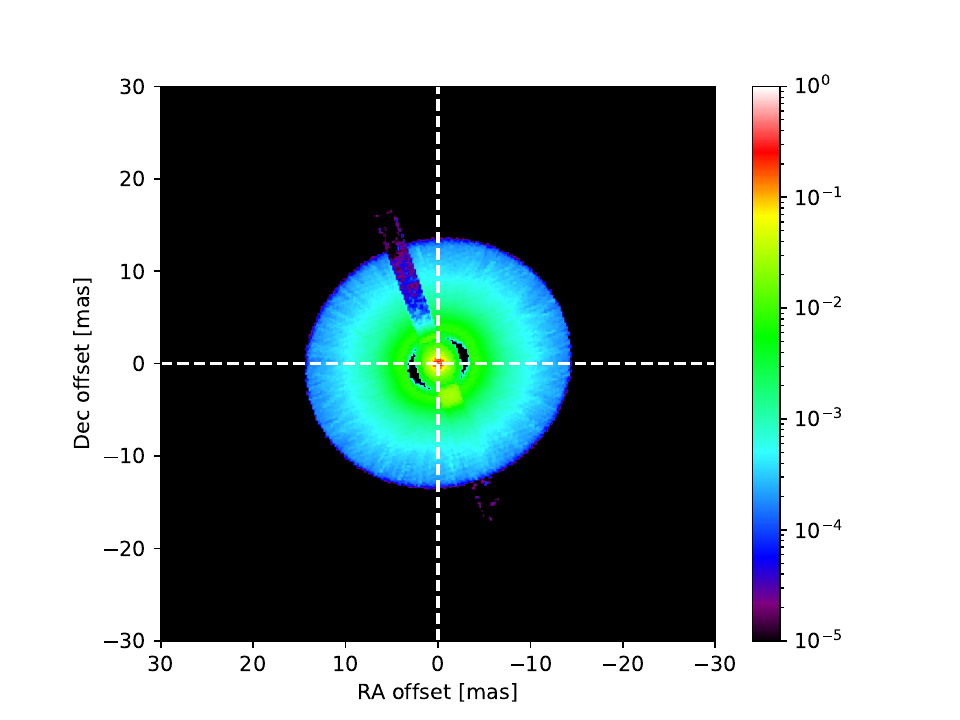}
    \caption{$K$ band model image with a torus outer radius of 12 au.}
    \label{3to12image}
\end{figure}

\begin{figure}%[hbt!]
    \centering
    \includegraphics[clip, width=\columnwidth]{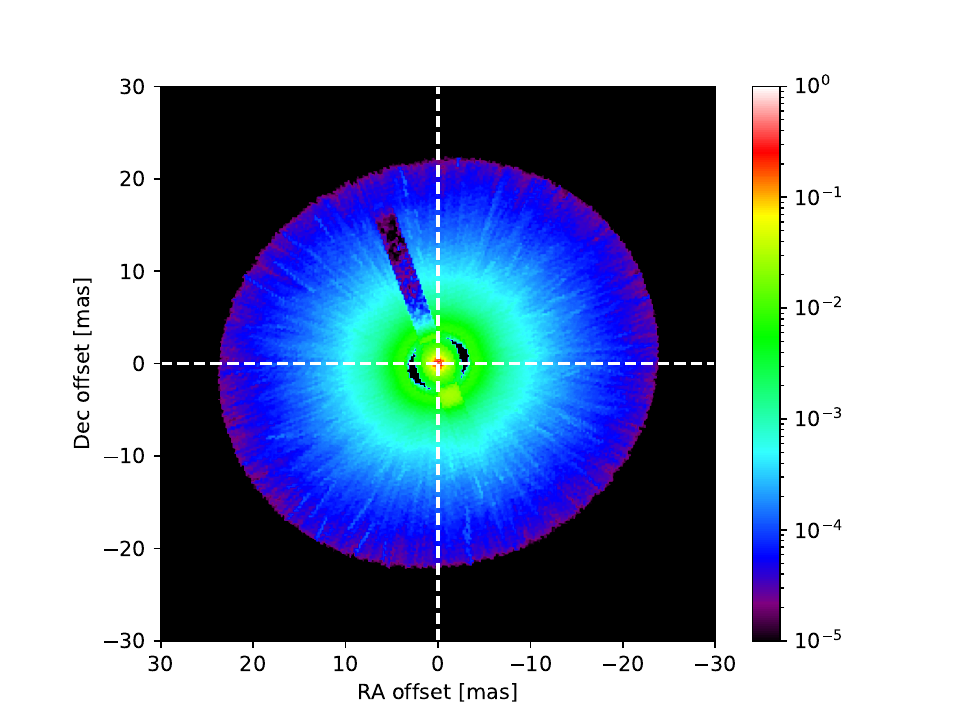}
    \caption{$K$ band model image with a torus outer radius of 20 au.}
    \label{3to20image}
\end{figure}

\begin{figure}%[hbt!]
    \centering
    \includegraphics[clip, width=\columnwidth]{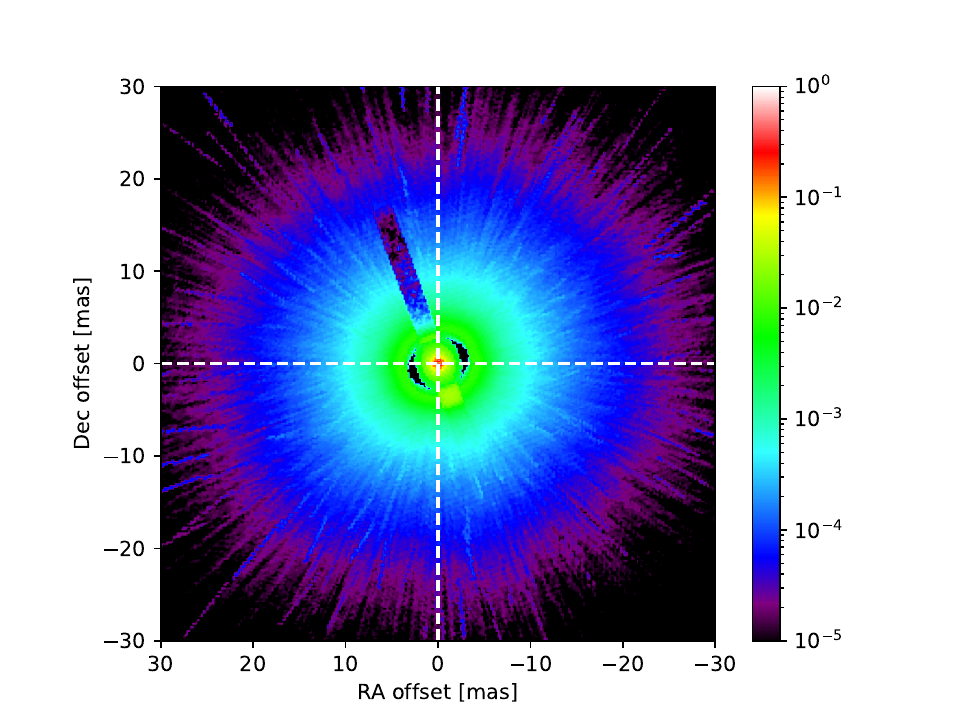}
    \caption{$K$ band model image with a torus outer radius of 40 au.}
    \label{3to40image}
\end{figure}

\begin{figure}%[hbt!]
    \centering
    \includegraphics[clip, width=\columnwidth]{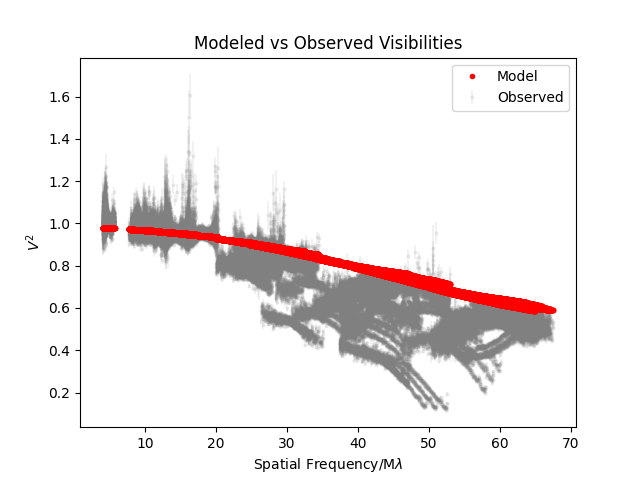}
    \caption{Squared visibilities for the $K$ band model image with an outer radius of 8 au.}
    \label{3to8vis2}
\end{figure}

\begin{figure}%[hbt!]
    \centering
    \includegraphics[clip, width=\columnwidth]{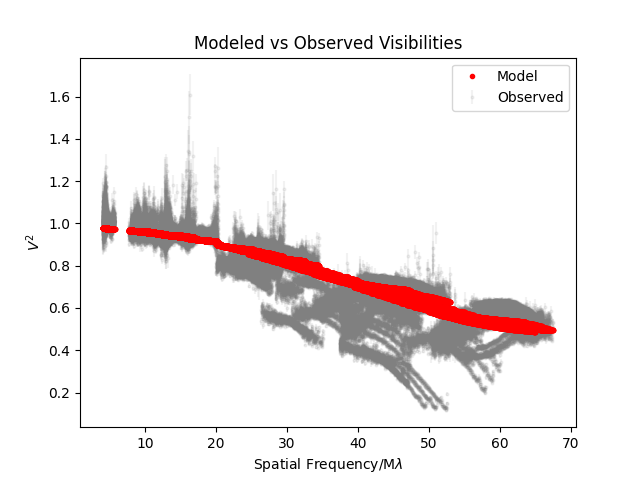}
    \caption{Squared visibilities for the $K$ band model image with a torus outer radius of 12 au.}
    \label{3to12vis2}
\end{figure}

\begin{figure}%[hbt!]
    \centering
    \includegraphics[clip, width=\columnwidth]{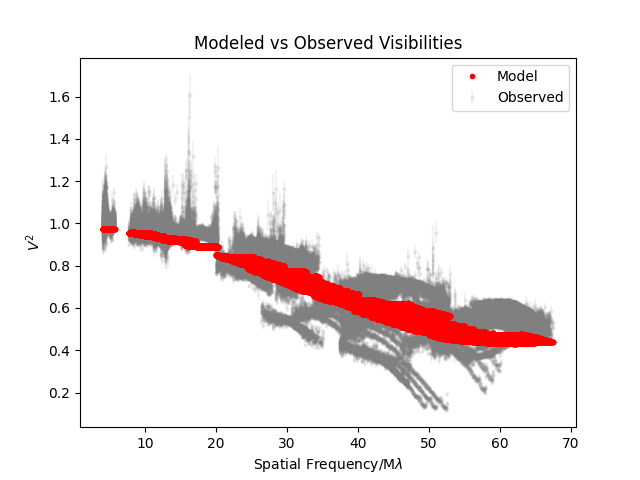}
    \caption{Squared visibilities for the $K$ band model image with a torus outer radius of 20 au.}
    \label{3to20vis2}
\end{figure}

\begin{figure}%[hbt!]
    \centering
    \includegraphics[clip, width=\columnwidth]{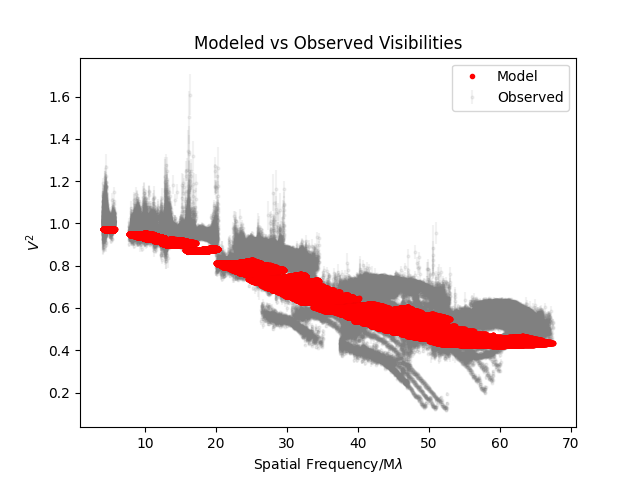}
    \caption{Squared visibilities for the $K$ band model image with a torus outer radius of 40 au.}
    \label{3to40vis2}
\end{figure}

\clearpage
\newpage

%######## CLOSURE PHASES

\section{Additional $H$ band model} \label{appendix-H}

\begin{figure}%[hbt!]
    \centering
    \includegraphics[clip, width=\columnwidth]{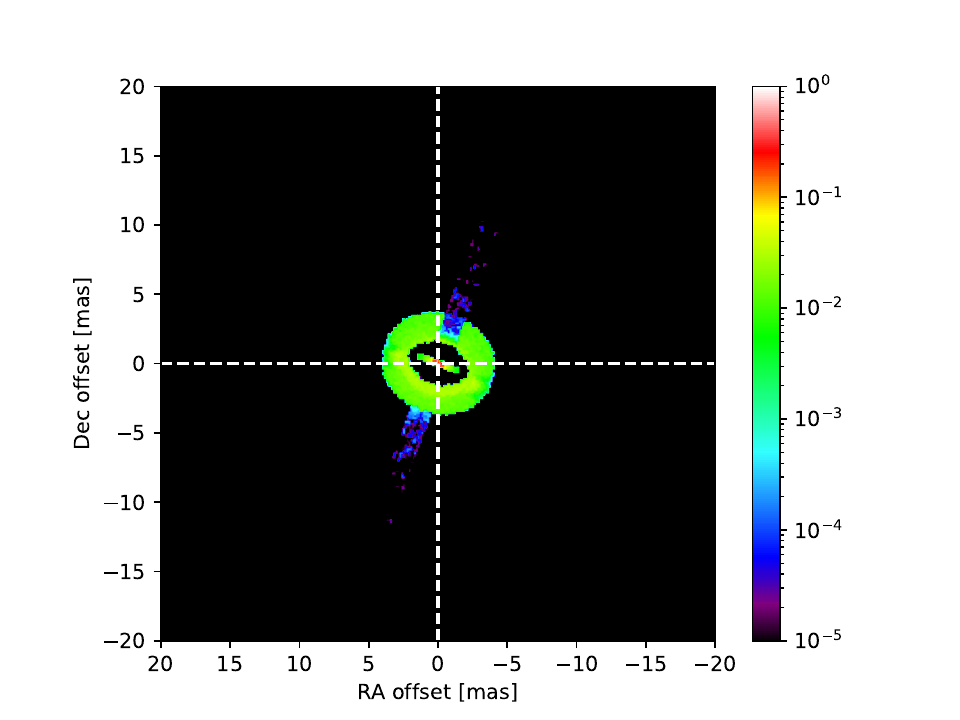}
    \caption{One of the many additional $H$ band models that were constructed for the purpose of explaining the observed $H$ band squared visibilities (see text).}
    \label{additionalHband}
\end{figure}

\begin{figure}%[hbt!]
    \centering
    \includegraphics[clip, width=\columnwidth]{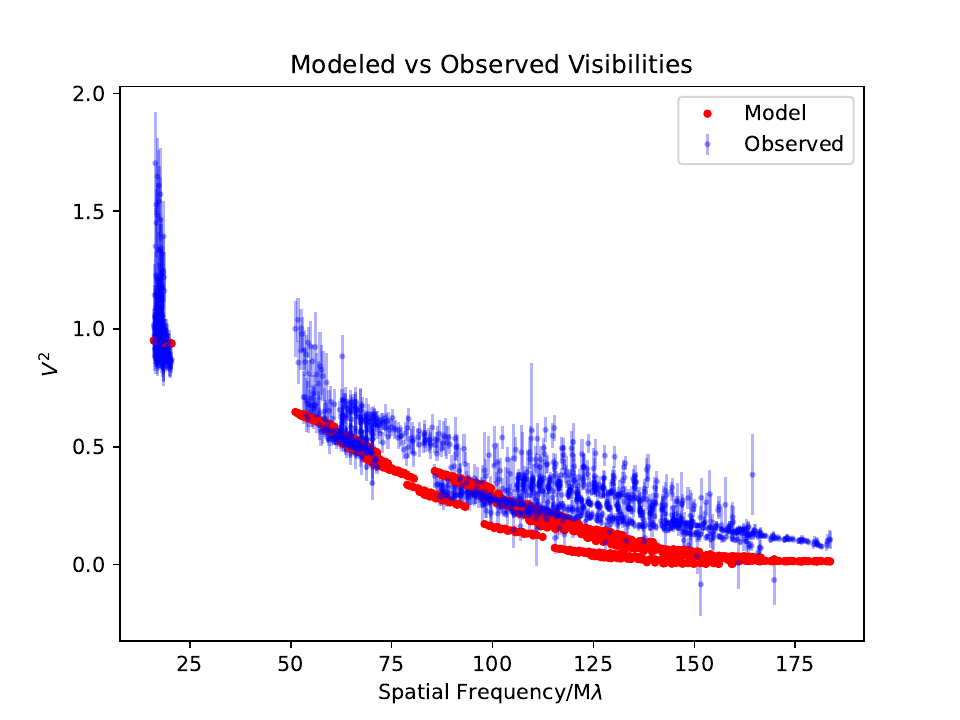}
    \caption{Simulated $H$ band squared visibilities for the extra model (see text).}
    \label{additionalHvis2}
\end{figure}

\begin{figure}%[hbt!]
    \centering
    \includegraphics[clip, width=\columnwidth]{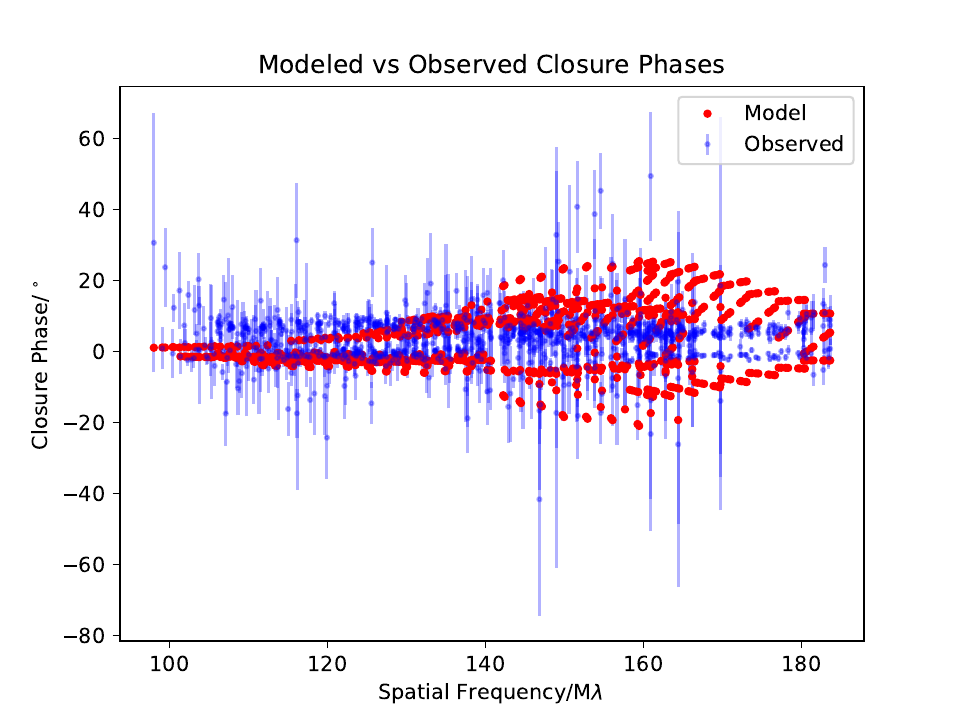}
    \caption{Simulated $H$ band closure phases for the extra model.}
    \label{additionalHt3}
\end{figure}

\section{Models with different orientations} \label{appendix-angles}

\begin{figure}%[hbt!]
    \centering
    \includegraphics[clip, width=\columnwidth]{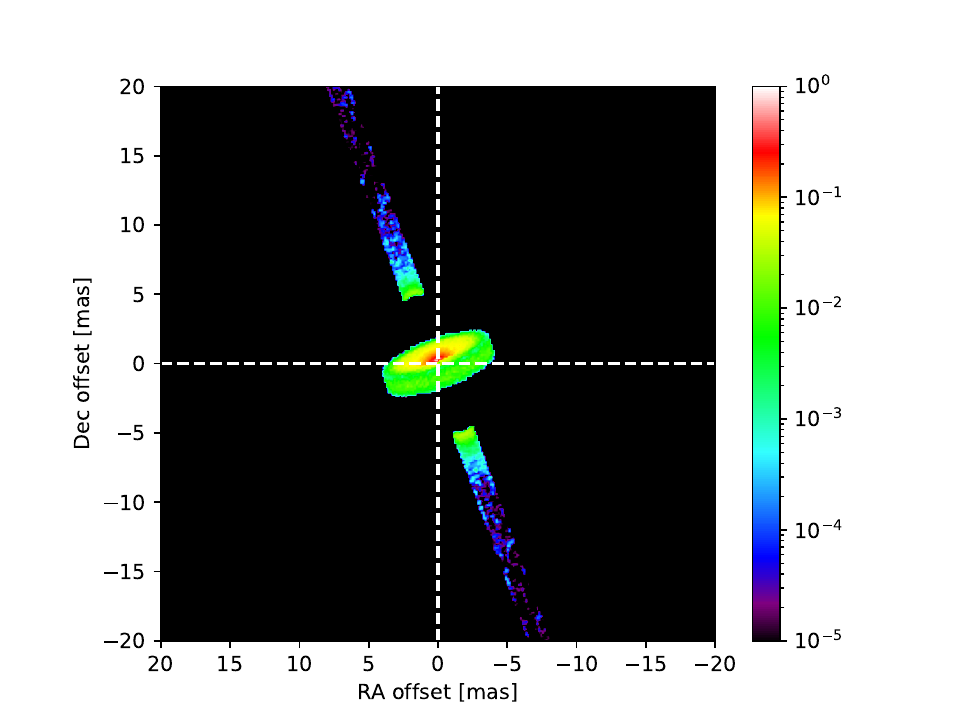}
    \caption{$H$ band model image with $i_{obs} =50\degree$. The colorbar shows normalized logarithmic intensity.}
    \label{incl50}
\end{figure}

\begin{figure}%[hbt!]
    \centering
    \includegraphics[clip, width=\columnwidth]{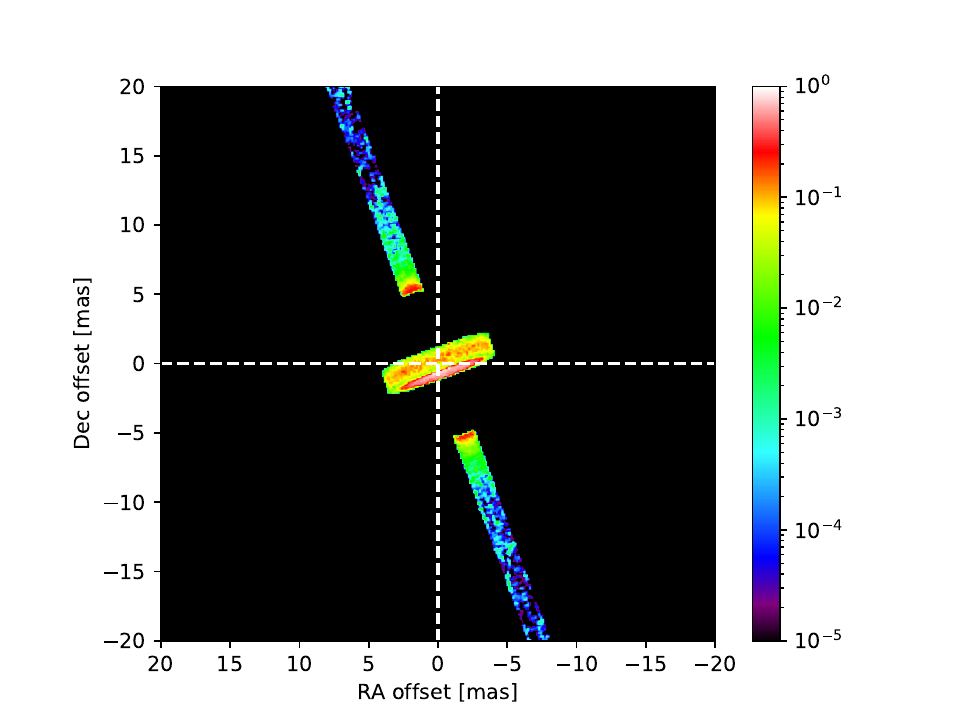}
    \caption{$H$ band model image with $i_{obs} =70\degree$. The colorbar shows normalized logarithmic intensity.}
    \label{incl70}
\end{figure}

\begin{figure}%[hbt!]
    \centering
    \includegraphics[clip, width=\columnwidth]{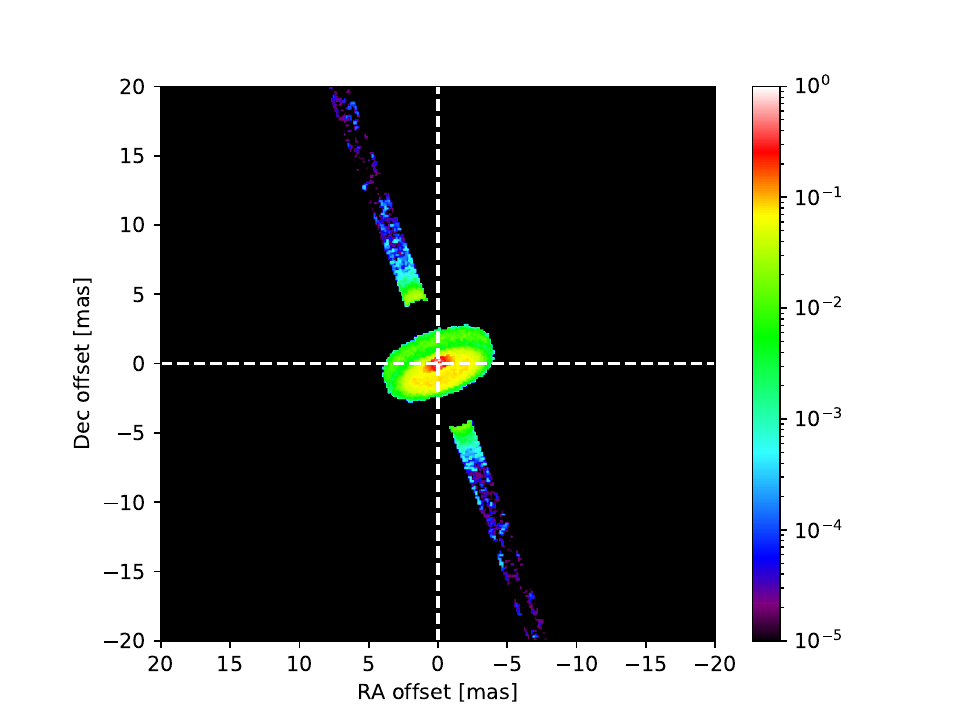}
    \caption{$H$ band model image with $i_{obs} =90\degree$. The colorbar shows normalized logarithmic intensity.}
    \label{incl90}
\end{figure}

\begin{figure}%[hbt!]
    \centering
    \includegraphics[clip, width=\columnwidth]{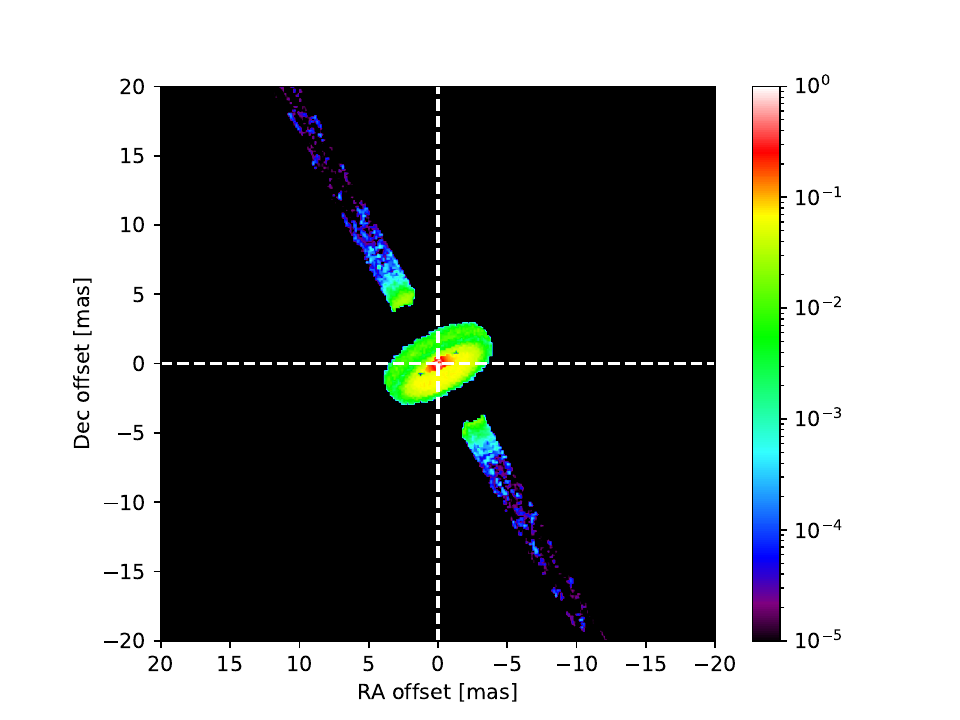}
    \caption{$H$ band model image with $i_{obs} =90\degree$ and $\phi_{obs}=20\degree$. The colorbar shows normalized logarithmic intensity.}
    \label{incl90phi20}
\end{figure}

\begin{figure}%[hbt!]
    \centering
    \includegraphics[clip, width=\columnwidth]{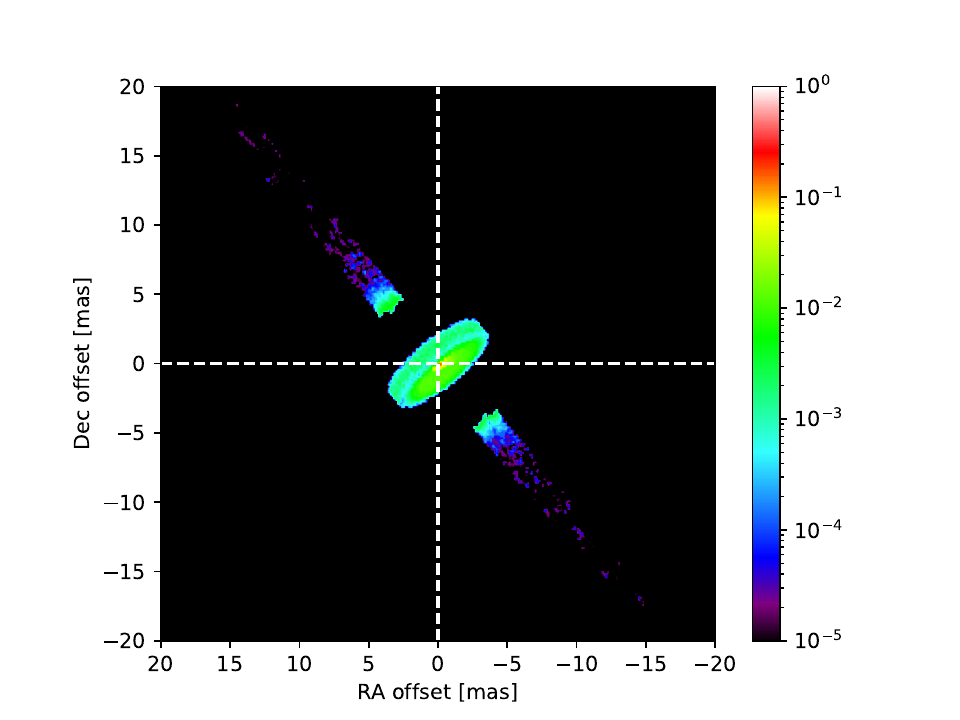}
    \caption{$H$ band model image with $i_{obs} =90\degree$ and $\phi_{obs}=50\degree$. The colorbar shows normalized logarithmic intensity.}
    \label{incl90phi50}
\end{figure}

\begin{figure}%[hbt!]
    \centering
    \includegraphics[clip, width=\columnwidth]{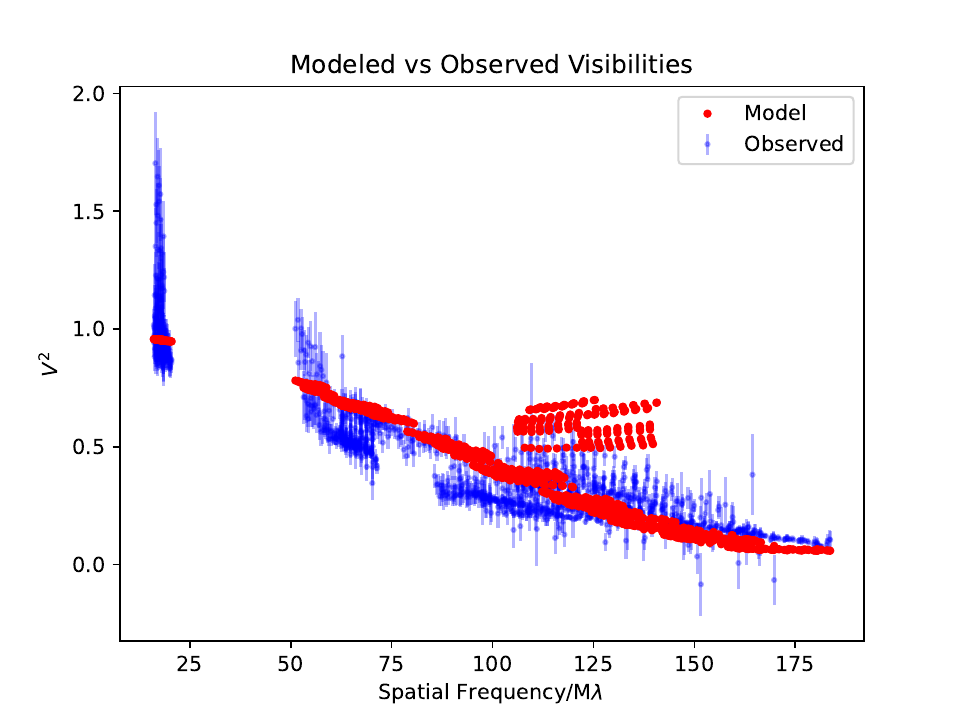}
    \caption{$H$ band model (red) squared visibilities for the $i_{obs} =50\degree$ image and observations (blue).}
    \label{incl50vis2}
\end{figure}

\begin{figure}%[hbt!]
    \centering
    \includegraphics[clip, width=\columnwidth]{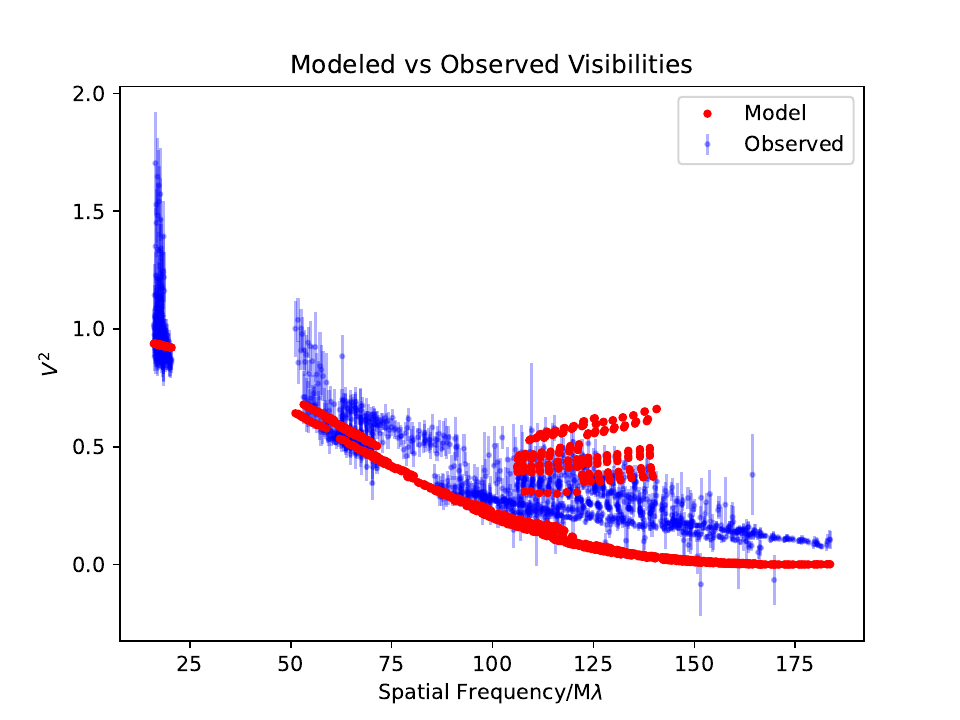}
    \caption{$H$ band model (red) squared visibilities for the $i_{obs} =70\degree$ image and observations (blue).}
    \label{incl70vis2}
\end{figure}

\begin{figure}%[hbt!]
    \centering
    \includegraphics[clip, width=\columnwidth]{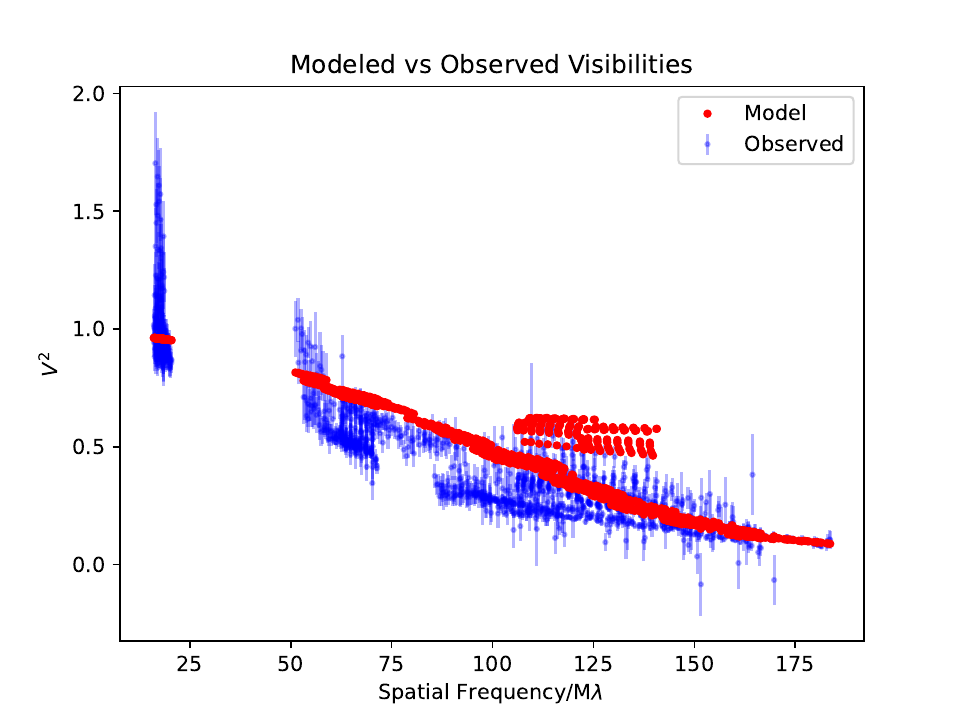}
    \caption{$H$ band model (red) squared visibilities for the $i_{obs} =90\degree$ image and observations (blue).}
    \label{incl90vis2}
\end{figure}

\begin{figure}%[hbt!]
    \centering
    \includegraphics[clip, width=\columnwidth]{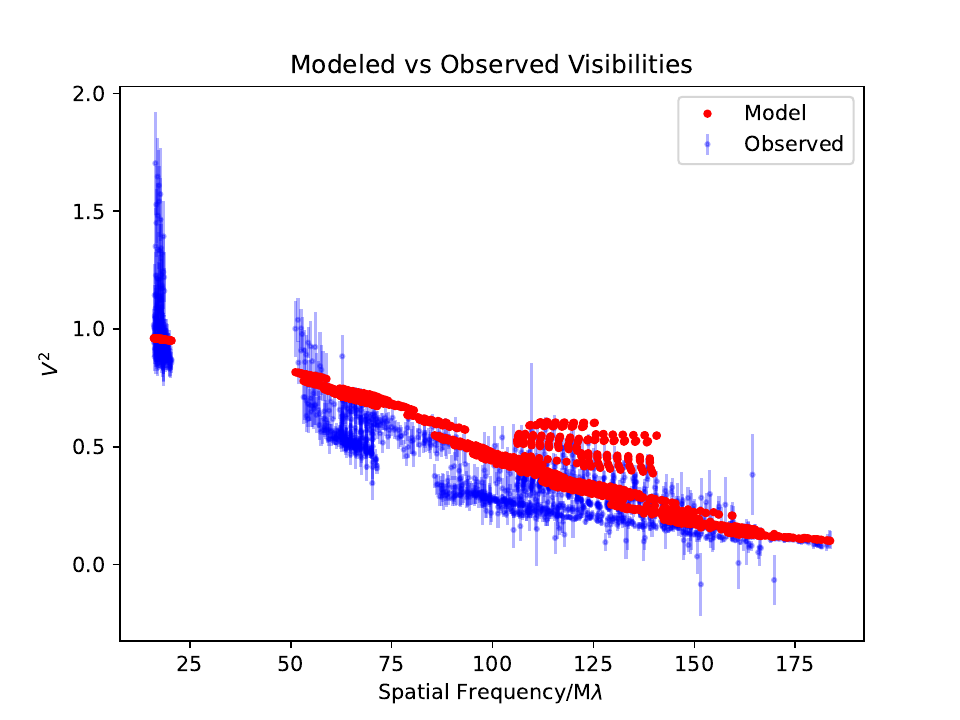}
    \caption{$H$ band model (red) squared visibilities for the $i_{obs} =90\degree$, $\phi_{obs}=20\degree$ image and observations (blue).}
    \label{incl90phi20vis2}
\end{figure}

\begin{figure}%[hbt!]
    \centering
    \includegraphics[clip, width=\columnwidth]{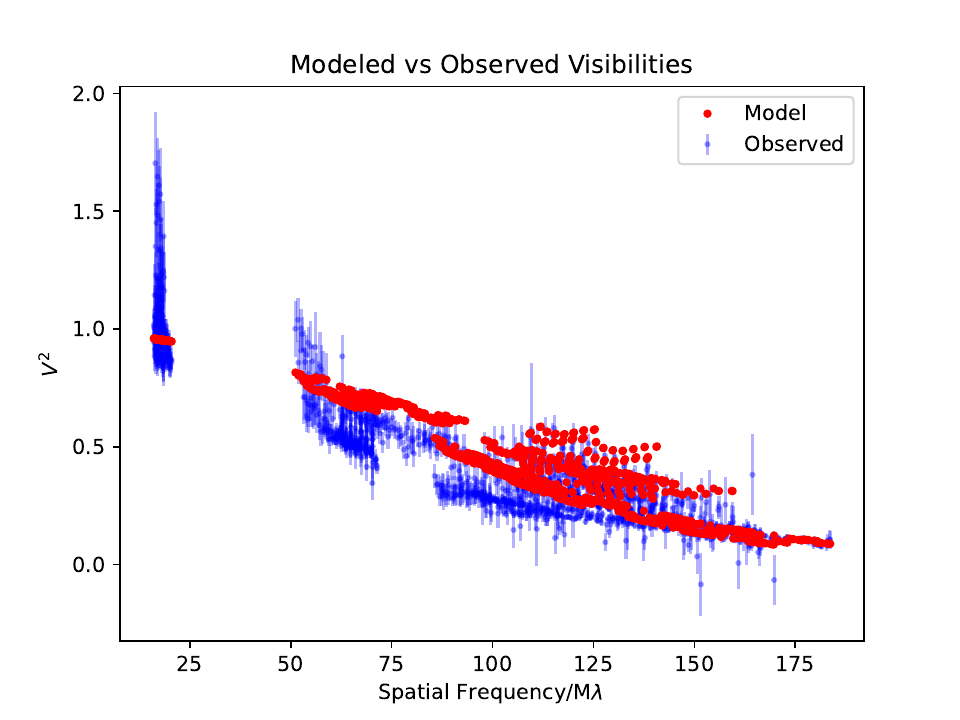}
    \caption{$H$ band model (red) squared visibilities for the $i_{obs} =90\degree$, $\phi_{obs}=50\degree$ image and observations (blue).}
    \label{incl90phi50vis2}
\end{figure}

\begin{figure}%[hbt!]
    \centering
    \includegraphics[clip, width=\columnwidth]{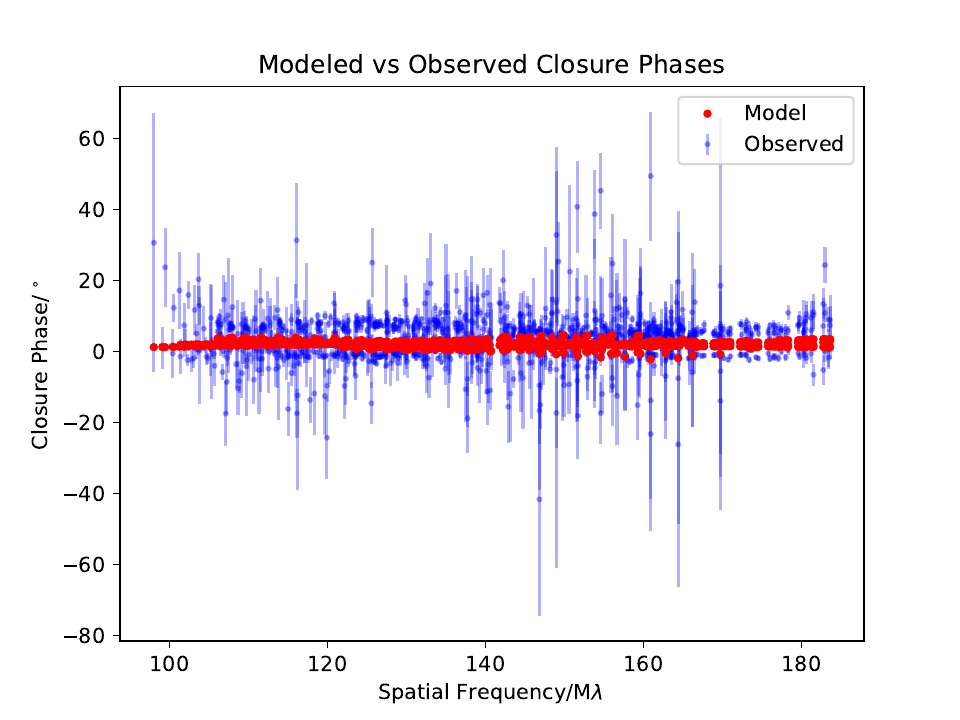}
    \caption{$H$ band model (red) closure phases for the $i_{obs} =50\degree$ image and observations (blue).}
    \label{incl50t3}
\end{figure}

\begin{figure}%[hbt!]
    \centering
    \includegraphics[clip, width=\columnwidth]{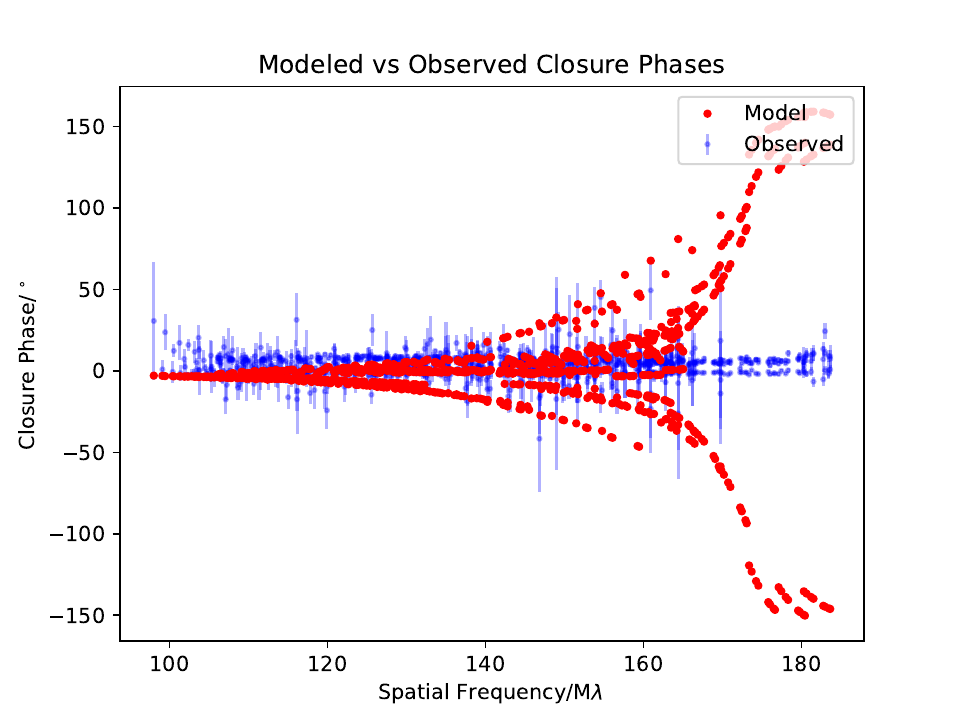}
    \caption{$H$ band model (red) closure phases for the $i_{obs} =70\degree$ image and observations (blue).}
    \label{incl70t3}
\end{figure}

\begin{figure}%[hbt!]
    \centering
    \includegraphics[clip, width=\columnwidth]{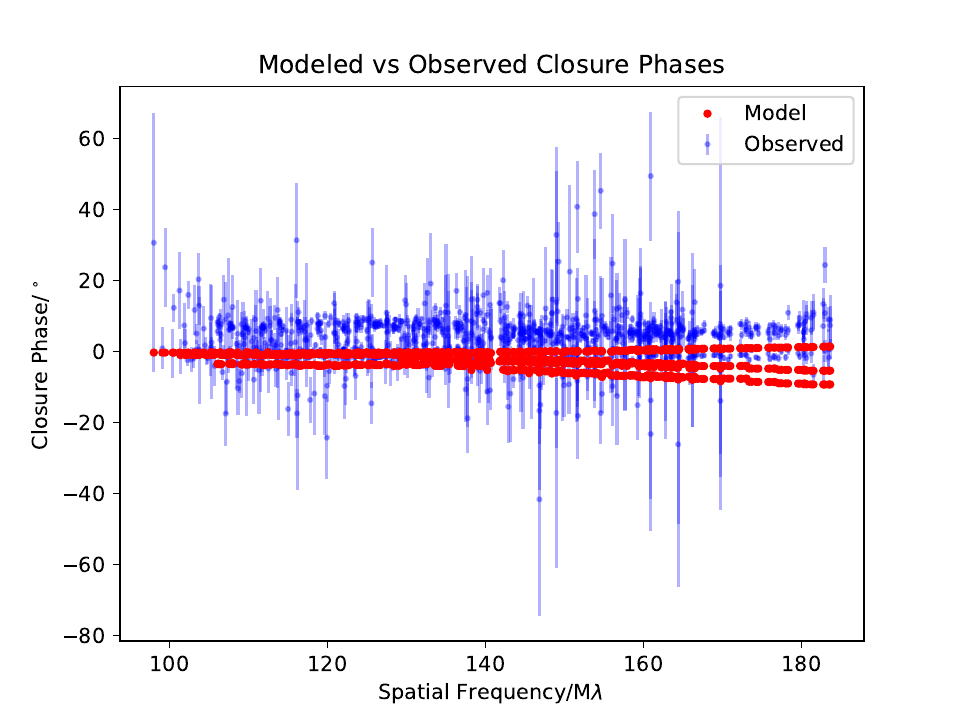}
    \caption{$H$ band model (red) closure phases for the $i_{obs} =90\degree$ image and observations (blue).}
    \label{incl90t3}
\end{figure}

\begin{figure}%[hbt!]
    \centering
    \includegraphics[clip, width=\columnwidth]{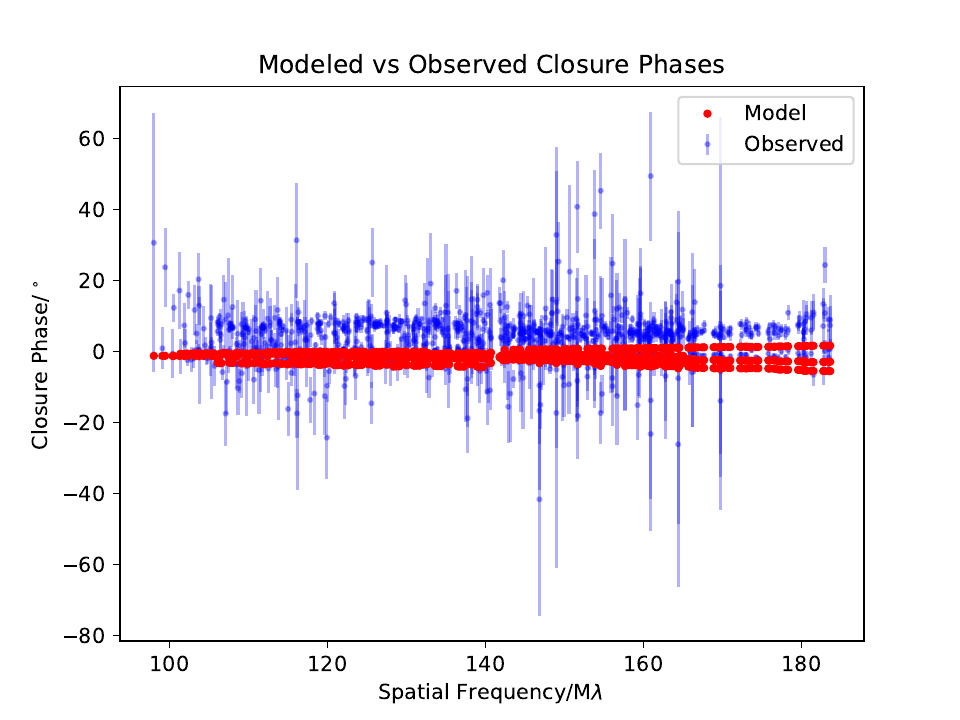}
    \caption{$H$ band model (red) closure phases for the $i_{obs} =90\degree$ and $\phi_{obs}=20\degree$ image and observations (blue).}
    \label{incl90phi20t3}
\end{figure}

\begin{figure}%[hbt!]
    \centering
    \includegraphics[clip, width=\columnwidth]{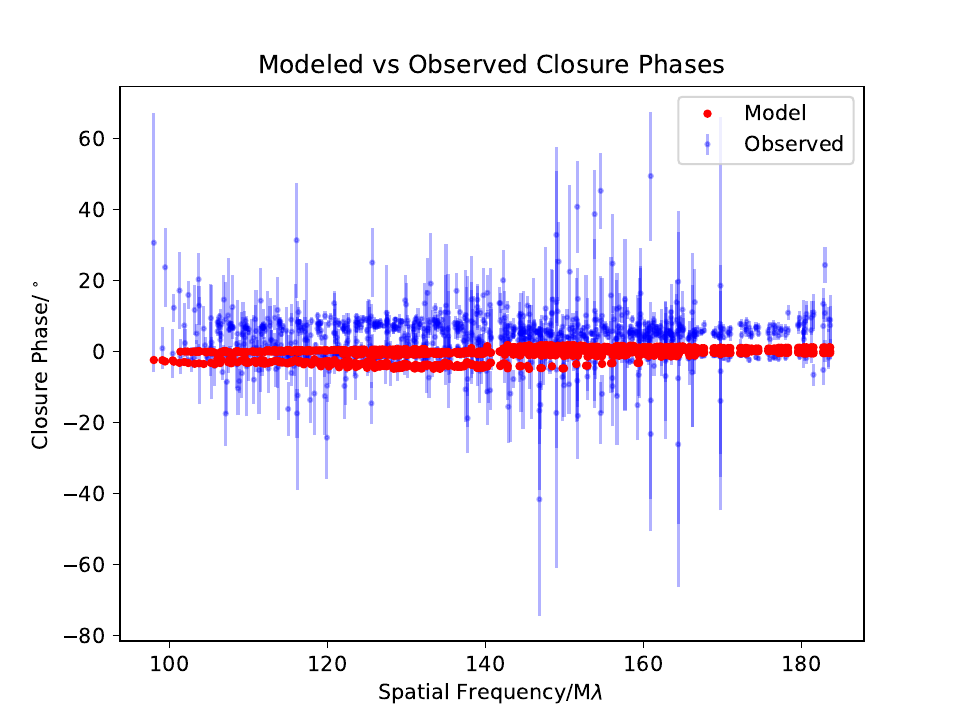}
    \caption{$H$ band model (red) closure phases for the $i_{obs} =90\degree$ and $\phi_{obs}=50\degree$ image and observations (blue).}
    \label{incl90phi50t3}
\end{figure}

\clearpage
\newpage
\section{Extra polarization maps} \label{appendix-polarization-maps}

% and \ref{hbandpol}. 

\begin{figure}%[hbt!]
    \centering
    \includegraphics[trim=10 5 0 20,clip, width=\columnwidth]{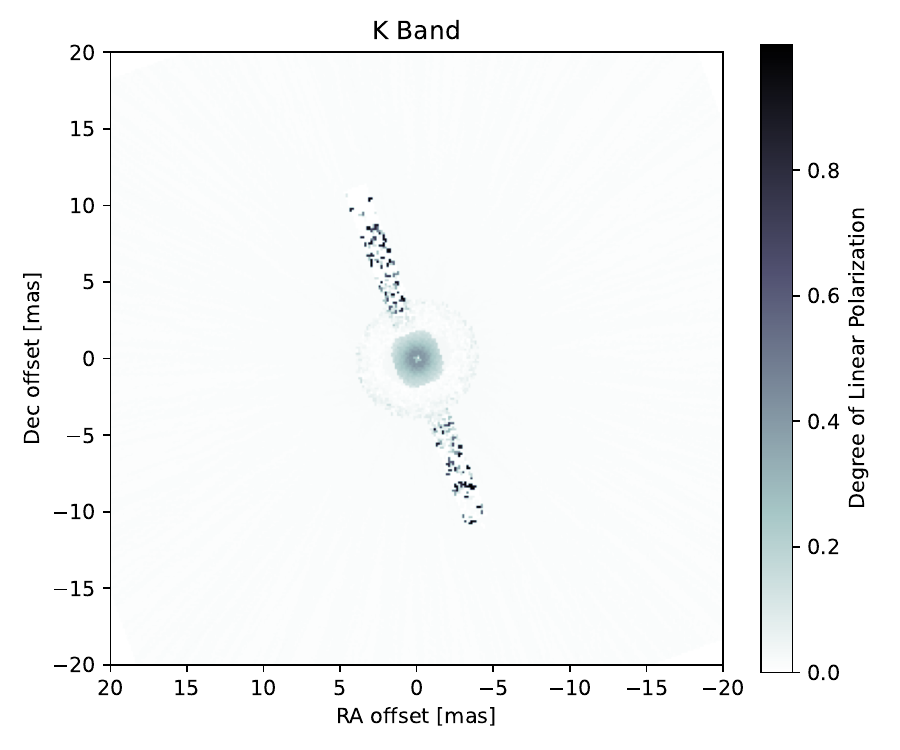}
%    \caption{$K$ band simulated image with degree of linear polarization.}
%    \label{kbandpol}
%\end{figure}
%\begin{figure}%[hbt!]
%    \centering
    \includegraphics[trim=10 5 0 20,clip, width=\columnwidth]{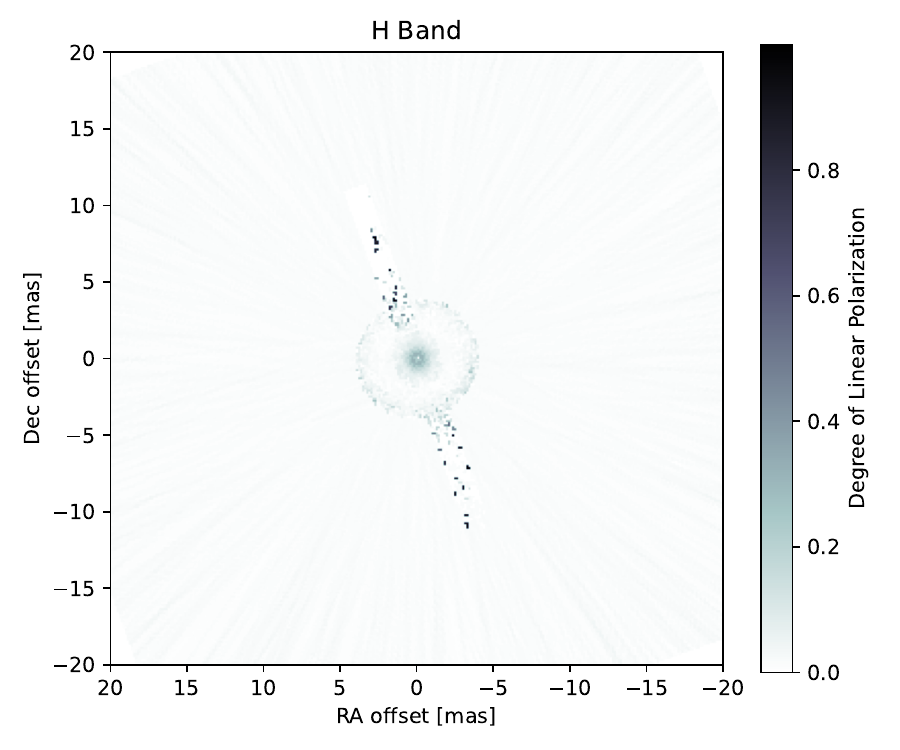}
    \caption{Simulated maps of degree of linear polarization in $K$ (top) and $H$ (bottom) bands.}
    \label{khbandpol}
\end{figure}

\begin{figure}%[hbt!]
    \centering
    \includegraphics[clip, width=\columnwidth]{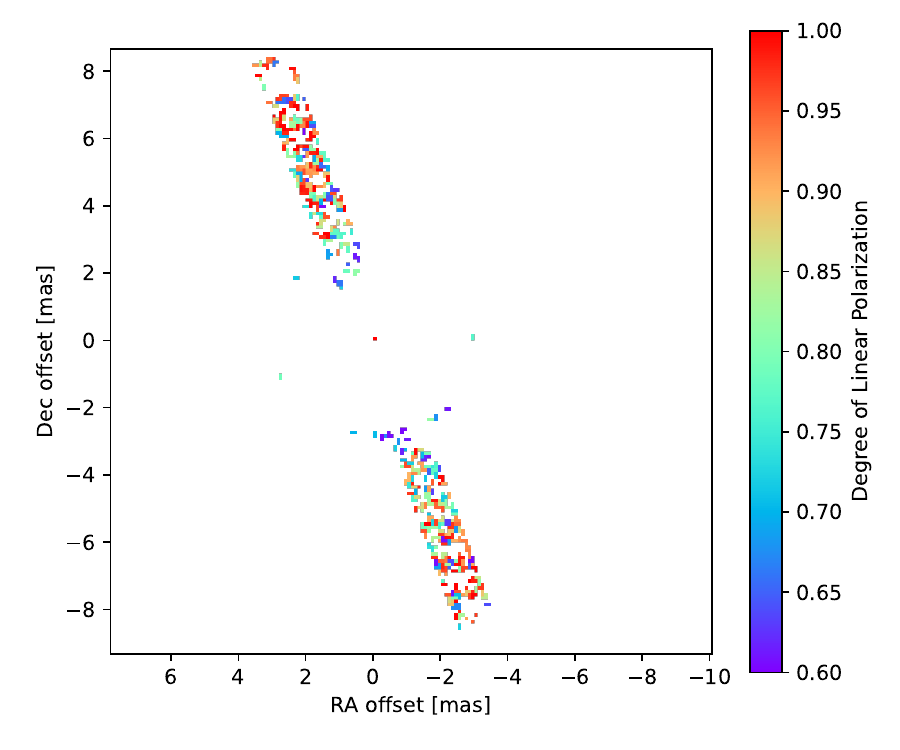}
    \includegraphics[clip, width=\columnwidth]{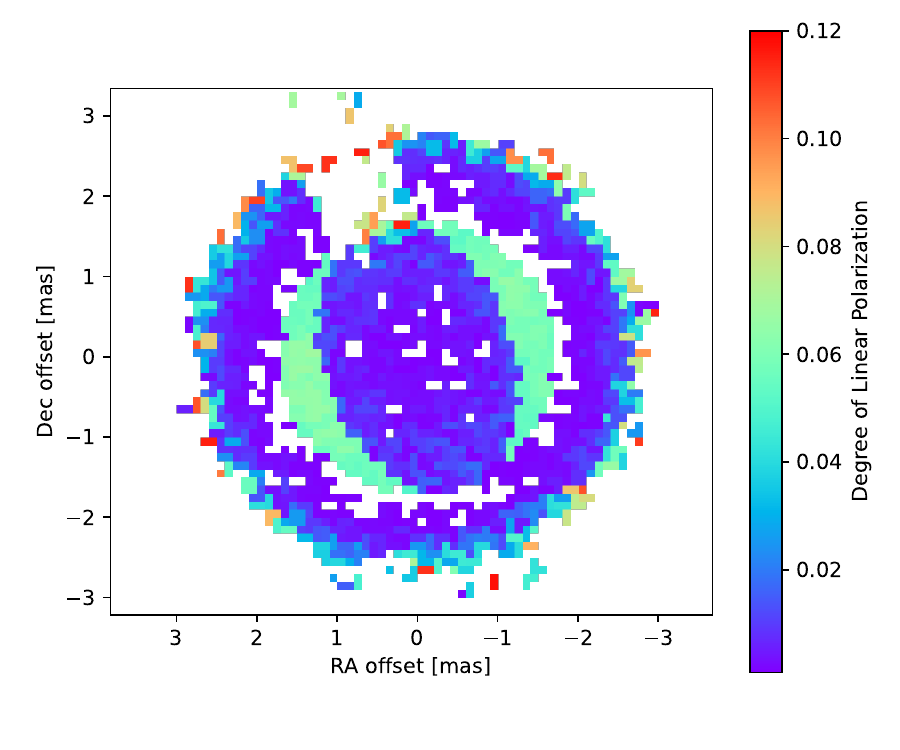}
    \caption{Simulated maps of degree of linear polarization (DoLP) in the $V$ band, with intensity scales optimised for the jets (top) and the torus (bottom). The maps have been zoomed in to show the degree of polarization across both structures. From the maps, we can see that the jets are strongly polarized with DoLP values $>0.5$, while the torus is weakly polarized with DoLP values $<0.2$.} 
    \label{polmapsboth}
\end{figure}

\end{appendix}

\end{document}